\documentclass[a4paper,11pt]{article}

\pdfoutput=1

\usepackage{jcappub}

\usepackage[T1]{fontenc}
\usepackage{blindtext}
\usepackage[utf8]{inputenc}
\usepackage{graphicx}
\usepackage{subfigure}
\usepackage{dcolumn}
\usepackage{bm}
\usepackage{accents}
\usepackage{epsfig}
\usepackage{amsfonts}
\usepackage{amsmath}
\usepackage{amssymb}
\usepackage[usenames]{color}
\usepackage[dvipsnames]{xcolor}
\usepackage{float}
\usepackage{caption}
\usepackage{booktabs}
\usepackage{arydshln}
\usepackage{multirow}
\usepackage{multicol}
\usepackage{tensor}
\usepackage{titlesec}
\usepackage{hyperref}
\usepackage[shortlabels]{enumitem}
\usepackage[normalem]{ulem}
\usepackage{fixltx2e}

\usepackage{seqsplit}
\newcommand{\hash}[1]{{\ttfamily\seqsplit{#1}}}

\graphicspath{./figures/}

\numberwithin{equation}{section}

\DeclareMathOperator{\sech}{sech}

\begin{document}

\title{\boldmath \huge Moduli Dynamics in Effective Nested Warped Geometry in Four Dimensions and Some Cosmological Implications}



\author[*,1]{Arko Bhaumik \note{Corresponding author}}
\author[\dagger]{and Soumitra SenGupta}

\affiliation[*]{Physics and Applied Mathematics Unit, Indian Statistical Institute, \\ 203, B.T. Road, Kolkata 700 108, India}
\affiliation[\dagger]{School of Physical Sciences, Indian Association for the Cultivation of Science, \\ 2A \& 2B, Raja Subodh Chandra Mallick Road, Kolkata 700 032, India}

\emailAdd{arkobhaumik12@gmail.com}
\emailAdd{tpssg@iacs.res.in}

\abstract{We analyze the effective four-dimensional dynamics of the extra-dimensional moduli fields in curved braneworlds having nested warping, with particular emphasis on the doubly warped model which is interesting in the light of current collider constraints on the mass of the Kaluza-Klein graviton. The presence of a non-zero brane cosmological constant ($\Omega$) naturally induces an effective moduli potential in the four-dimensional action, which shows distinct features in dS ($\Omega>0$) and AdS ($\Omega<0$) branches. For the observationally interesting case of dS 4-branes, a metastable minimum in the potential arises along the first modulus, with no minima along the higher moduli. The underlying nested geometry also leads to interesting separable forms of the non-canonical kinetic terms in the Einstein frame, where the brane curvature directly impacts the kinetic properties of only the first modulus. The non-canonicity of the scenario has been illustrated via an explicit computation of the field space curvature. We subsequently explore the ability of curved multiply warped geometries to drive inflation with an in-built exit mechanism, by considering predominant slow roll along each modular direction on a case-by-case basis. We find slow roll on top of the metastable plateau along the first modular direction to be the most viable scenario, with the higher-dimensional moduli parametrically tuning the height of the potential without significant impact on the inflationary observables. On the other hand, while slow roll along the higher moduli can successfully inflate the background and eventually lead to an exit, consistency with observations seemingly requires unphysical hierarchies among the extra-dimensional radii, thus disfavouring such scenarios.}

\maketitle

\section{Introduction} \label{sec:intro}

Among the wide range of phenomenological extra-dimensional theories, warped braneworld models have occupied a prominent spot for over two decades. Within this class of scenarios, the prototypical five-dimensional (5D) Randall-Sundrum model (RS1) \cite{Randall:1999ee} exhibits a non-factorizable topology of the form $\mathcal{M}(1,3)\times S^1/\mathbb{Z}_2$, where an AdS$_5$ bulk with a large cosmological constant induces exponential warping along the compact extra dimensional orbifold of radius $r_c\sim M_5^{-1}$, with $M_5$ being the fundamental Planck scale. Consequently, a large hierarchy may develop between the physical mass scales of the two 3-branes situated at the opposite poles of the orbifold, which comes naturally without any extreme fine tuning of parameters. Such a model was famously shown capable of explaining the gauge hierarchy of the Standard Model (SM) in an elegant manner, while keeping the four-dimensional effective Planck scale, $M_{\textrm{Pl}}$, close to $M_5$ so as to avoid any new hierarchy. Augmented with an appropriate stabilization mechanism for the extra-dimensional modulus \cite{Goldberger:1999uk,Brevik:2000vt,Dey:2006px,Chakraborty:2016gpg,Das:2017htt,Elahi:2022hpj}, the RS1 model subsequently offered a wealth of phenomenological predictions at the electroweak scale \cite{Goldberger:1999un,Davoudiasl:1999tf,Davoudiasl:1999jd,DeWolfe:1999cp,Chang:1999nh,Csaki:2000zn,Huber:2000ie,Agashe:2003zs,Luty:2004ye,Davoudiasl:2005uu,Chacko:2013dra,Ahmed:2019zxm,Lee:2021wau,Frank:2023lxf}. But the continued null detection of an $\mathcal{O}$(TeV) massive Kaluza-Klein (KK) graviton mode at collider experiments has considerably narrowed the parameter space of the original 5D proposal \cite{ATLAS:2011ab,ATLAS:2012hvw,ATLAS:2014pcp,CMS:2014mws,ATLAS:2015shg,CMS:2015cwa,CMS:2016crm,ATLAS:2017zuf,ATLAS:2018rnh,ATLAS:2019erb}. To evade these bounds, a small hierarchy needs to be considered between the vacuum expectation value (VEV) of the fundamental Higgs scalar, $v_H$, and the 5D Planck scale, with a magnitude of $v_H/M_5\lesssim10^{-2}$ \cite{Arun:2014dga}. In principle, this allows the breakdown of the effective four-dimensional (4D) theory resulting from RS1 and the possibility of new physics at least two orders of magnitude below $M_5$. Unless the cut-off of the 4D effective theory, which is furnished by $r_c^{-1}$, is also lowered by a similar amount, the situation entails fine tuning of at least 2$-$3 orders of magnitude in order to consistently accommodate such a reduced value of $v_H$. On the other hand, it is not easy to change the value of $r_c^{-1}$ so drastically due to the exponential dependence of the warp factor on the value of the modulus. This lands one right in the midst of the aforementioned fine tuning issue and spoils the elegance of the original RS1 model, whose principal \textit{forte} was to resolve the hierarchy problem of the Higgs mass without fine tuning of the fundamental parameters. 

Keeping these experimental constraints in sight, the prospects turn out to be significantly better for a six-dimensional analogue of RS1 with topology $\left[\mathcal{M}(1,3)\times S^1/\mathbb{Z}_2\right]\times S^1/\mathbb{Z}_2$, \textit{\textit{i.e.}} having two successive levels of ``nested'' warping along two orbifolds \cite{Choudhury:2006nj}. This system, which is a particular one among several proposed higher-dimensional extensions of the RS1 idea \cite{Randjbar-Daemi:2000bjr,PhysRevD.64.044021,PhysRevLett.90.101601,Kaloper:2004cy,PhysRevD.72.064008,PhysRevD.77.124046,Archer:2010bm,Feng:2015sfa,Meiers:2017ltj,Wan:2020smy}, consists of four 4-branes acting as ``walls'' of an AdS$_6$ bulk, with four 3-branes located at the ``corners'' where the 4-branes intersect pairwise. Depending on the two extra-dimensional moduli, these corner branes may be unequally warped. The geometric structure leads to a viable scenario where two of these branes are clustered close to the Planck scale, and the other two are clustered around the TeV scale, with a minor hierarchy developing within each pair. Such a set-up leads to a heavier mass as well as reduced coupling strengths to SM fields for the first KK graviton mode, which help evade current collider constraints without recourse to any ``little'' hierarchy unlike RS1 \cite{Arun:2014dga,Arun:2015ubr}. Besides this advantage, the doubly warped scenario offers other interesting phenomenological possibilities, a notable one being a geometric explanation of the SM fermion mass hierarchy \cite{Choudhury:2006nj,Hundi:2011dc}. This can be achieved by treating the SM fermions as 5D fields percolating into the bulk, with the brane-localized boundary kinetic terms altering the 4D effective scalar-fermion Yukawa couplings and leading to a minor splitting among the fermion masses. Further aspects of this 6D model have been studied in a series of subsequent works, pertaining to both moduli stabilization and radion dynamics \cite{Arun:2016csq,Bhaumik:2022xtd}, as well as the phenomenology of bulk gauge and matter fields \cite{Das:2011fb,Chakraborty:2014xda,Arun:2015kva,Arun:2016ela,Barman:2022qix}.

On the other hand, the negative tension of the visible brane is a disconcerting feature of RS1 due to stability issues within both Einstein gravity and its extensions \cite{Shiromizu:1999wj,Ida:2001qw,Charmousis:2003sq}. A particularly simple way around this problem is to consider globally curved branes, which makes it possible for both the branes to have positive tensions \cite{Das:2007qn}. This curvature originates from a 4D effective cosmological constant induced on the 3-branes, which may lead to either an anti-de Sitter (AdS) or a de Sitter (dS) geometry, with the latter one more interesting from a cosmological perspective. For such a simple modification, the new model in fact has quite a lot to offer. An exciting feature of the 5D non-flat scenario is the natural emergence of a 4D effective potential even in the absence of any bulk field, that vanishes identically in the flat limit. While it is possible to adapt the conventional Goldberger-Wise approach to the 5D non-flat case \cite{Koley:2008hs}, this induced potential itself can serve to stabilize the modulus in a metastable manner \cite{Banerjee:2017jyk}, thereby obviating the need for additional bulk fields to a considerable extent. Moreover, it can successfully lead to a non-canonical slow roll inflationary phase \cite{Banerjee:2018kcz} as well as a non-singular bouncing scenario \cite{Banerjee:2020uil}, both concordant with observations. These features serve to establish the fact that non-flat brane geometry in warped braneworld models is more than a mathematical curiosity, and is of direct relevance from an experimental angle as well as from the perspective of theoretical consistency.

Motivated by the reasons highlighted so far, we have generalized multiply warped braneworld scenarios to include non-zero brane curvature and explored their geometric properties in an earlier work \cite{Bhaumik:2023tmg}. In the present work, our aim is to study the resulting induced potential in multiply warped non-flat backgrounds, and to focus on some of its cosmological implications \textit{vis-\`{a}-vis} the singly warped non-flat model. In Sec. \ref{sec:review}, we start with a concise review of nested warped geometry in presence of non-zero brane curvature. We subsequently discuss the origin of the 4D induced potential in the curved doubly warped scenario in Sec. \ref{sec:modpot}, derive the effective non-canonical action for the moduli fields in the Einstein frame in terms of the Jordan frame action, and demonstrate the non-canonicity of the scenario under consideration by explicitly computing the curvature of the field space metric. In Sec. \ref{sec:inflation}, we investigate the scope of the resulting 4D effective framework in driving slow roll inflation while maintaining consistency with current observations. Moving on to Sec. \ref{sec:gentrends}, we highlight a few dynamical features which appear generic to non-flat geometries with more than two levels of nested warping. We conclude by summarizing our key results and discussing a few possible future directions in Sec. \ref{sec:disc}.

\section{Review of non-flat multiply warped scenario} \label{sec:review}

We start by briefly reviewing the key geometric aspects of the doubly warped non-flat model, which have been studied in detail in \cite{Bhaumik:2023tmg}. In the space-positive signature, the full 6D bulk-brane action can be written as
\begin{equation} \label{eq:fullact}
\begin{split}
\mathcal{S}=\int d^4x\int\limits_{-\pi}^{+\pi}dy\int\limits_{-\pi}^{+\pi}dz[&\sqrt{-g_6}\left(2M^4\mathcal{R}_6-\Lambda_6\right)-\sqrt{-g_5^{(\textrm{ind})}}\{V_1(z)\delta(y)+V_2(z)\delta(y-\pi)\}\\
& -\sqrt{-\bar{g}_5^{(\textrm{ind})}}\{V_3(y)\delta(z)+V_4(y)\delta(z-\pi)\}]\:,
\end{split}
\end{equation}
where $\mathcal{R}_6$ is the 6D Ricci scalar, $g_6$ is the 6D metric, $g_5^{(\textrm{ind})}$ ($\bar{g}_5^{(\textrm{ind})}$) is the induced 5D metric on the constant $y$-slices ($z$-slices), $V_i$ is the tension of the corresponding 4-brane whose location is denoted by the $\delta$-function, $\Lambda_6$ is the (negative) bulk cosmological constant, and $M$ is the fundamental (6D) Planck mass (identified hereafter with $M_6$ from the discussion in Sec. \ref{sec:intro}). The tension of each corner 3-brane is given by the pairwise algebraic sum of the two corresponding intersecting 4-branes. The general non-flat doubly warped metric ansatz associated with this action can be taken as 
\begin{equation}
ds^2=b(z)^2\left[a(y)^2g_{\mu\nu}dx^\mu dx^\nu+T_1^2dy^2\right]+T_2^2dz^2\:,
\end{equation}
where $a(y)$ and $b(z)$ are the two warp factors, and $T_1$ and $T_2$ are the orbifold radii of mass dimension $-1$. Note that the intrinsic 3-brane metric $g_{\mu\nu}$ is not identified \textit{a priori} with $\eta_{\mu\nu}$, \textit{\textit{i.e.}} one allows room for the possibility that the ``end of the world'' branes may be curved. This curvature can be interpreted as the result of a 4D cosmological constant, $\Omega$, induced geometrically on the corner branes, that non-trivially modifies the forms of both the warp factors and the brane tensions compared to the flat scenario. There are two possible branches depending on whether $\Omega<0$ or $\Omega>0$, which correspond to anti-de Sitter (AdS) and de Sitter (dS) geometries respectively. 
\subsection{Anti-de Sitter ($\Omega<0$)}
The solutions of the warp factors are given by
\begin{equation} \label{eq:warpfacsads}
a(y)=\omega_1\:\textrm{cosh}\left[\textrm{ln}\left(\dfrac{\omega_1}{c_1}\right)+c|y|\right]\:,\:\:b(z)=\dfrac{\textrm{cosh}(kz)}{\textrm{cosh}(k\pi)}\:,
\end{equation}
where $c$ and $k$ are the dimensionless moduli corresponding to the two orbifolds, with $y$ and $z$ being their respective angular coordinates. We define $\omega_1^2=-\dfrac{\Omega T_1^2}{3c^2}$ to be the dimensionless induced cosmological constant, and $c_1=1+\sqrt{1-\omega_1^2}$. The moduli and the radii are not all indepenent, but remain algebraically related via two equalities: $k=T_2\sqrt{-\dfrac{\Lambda_6}{40M^4}}$ and $c=\dfrac{T_1k}{T_2\textrm{cosh}(k\pi)}$. The 4-brane tensions, obtained by consistently matching boundary conditions, turn out to be
\begin{equation} \label{V11}
V_1(z)=24M^2\sqrt{-\dfrac{\Lambda_6}{40}\left(1-\omega_1^2\right)}\:\textrm{sech}(kz)\:,
\end{equation}
\begin{equation} \label{V12}
V_2(z)=24M^2\sqrt{-\dfrac{\Lambda_6}{40}}\:\textrm{tanh}\left(\textrm{ln}\dfrac{\omega_1}{c_1}+c\pi\right)\textrm{sech}(kz)\:,
\end{equation}
\begin{equation} \label{V13V14}
V_3(y)=0\:,\:\:V_4(y)=32M^2\sqrt{-\dfrac{\Lambda_6}{40}}\:\textrm{tanh}(k\pi)\:.
\end{equation}

\subsection{de Sitter ($\Omega>0$)}
For positive curvature, the solutions are given by
\begin{equation} \label{eq:warpfacsds}
a(y)=\omega_2\:\textrm{sinh}\left[\textrm{ln}\left(\dfrac{c_2}{\omega_2}\right)-c|y|\right]\:,\:\:b(z)=\dfrac{\textrm{cosh}(kz)}{\textrm{cosh}(k\pi)}\:,
\end{equation}
where $\omega_2^2=\dfrac{\Omega T_1^2}{3c^2}$ and $c_2=1+\sqrt{1+\omega_2^2}$, and the two other algebraic relations hold as earlier. The brane tensions in this regime are
\begin{equation} \label{V21}
V_1(z)=24M^2\sqrt{-\dfrac{\Lambda_6}{40}\left(1+\omega_2^2\right)}\:\textrm{sech}(kz)\:,
\end{equation}
\begin{equation} \label{V22}
V_2(z)=-24M^2\sqrt{-\dfrac{\Lambda_6}{40}}\textrm{coth}\left(\textrm{ln}\left(\dfrac{c_2}{\omega_2}\right)-c\pi\right)\textrm{sech}(kz)\:,
\end{equation}
\begin{equation} \label{V23V24}
V_3(y)=0\:,\:\:V_4(y)=32M^2\sqrt{-\dfrac{\Lambda_6}{40}}\textrm{tanh}(k\pi)\:.
\end{equation}
In both the regimes, it is easy to see that simultaneous large warping along both orbifolds, \textit{\textit{i.e.}} $c\sim k\gtrsim\mathcal{O}(1)$, necessitates a considerable hierarchy between $T_1$ and $T_2$, which is problematic from a model-building perspective in the sense that the scale of ``new physics'' would be significantly lowered from the fundamental Planck scale. The only way to resolve this issue is to demand unequal warping along the two orbifolds, so as to overall generate the $\mathcal{O}$(TeV) electroweak scale on a corner brane (preferably the maximally warped brane). This subsequently gives rise to two possible branches for each case. For AdS geometry, dominant warping along $y$ imposes a small upper bound $|\Omega|T_1^2\lesssim10^{-28}$ (which is interesting from a cosmological point of view) and can result in a maximally warped 3-brane with positive tension, whereas dominant warping along $z$ entails an extreme fine tuning of $\omega_1$ if it is to result in such a positive tension 3-brane while avoiding a large $T_1/T_2$ hierarchy. On the other hand, for dS geometry, positivity of the maximally warped 3-brane cannot be ensured, but dominant warping along $y$ places a similar upper bound on $\Omega$. Moreover, the tuning of $\Omega$ to its tiny observed value corresponds to the tuning of the extra dimensional moduli close to $M^{-1}$, which may lead to interesting features in the fermion mass hierarchy of the Standard Model (SM). These key results can be further generalized in a straightforward manner to $n$-fold warping alongside non-zero curvature. 

Finally, an interesting interpretation of the 6D model can be developed based on the fundamental brane tensions. Recall that the tension of the 4-brane situated at $z=0$ vanishes, \textit{\textit{i.e.}} $V_3(y)=0$, in both the AdS and the dS branches, as noted respectively in \eqref{V13V14} and \eqref{V23V24}. As such, in view of non-vanishing 4-brane tensions, it appears on physical grounds that the ``brane-box'' configuration with walls made up of the four 4-branes is effectively bounded along the $y$-direction by the two 4-branes at $y=0$ and at $y=\pi$, whereas along the $z$-direction it effectively consists of only one non-trivial 4-brane at $z=\pi$. We thus note that the $y$-direction mimics a 6D analogue of the RS1 scenario \cite{Randall:1999ee}, whereas the $z$-direction is similar to an RS2 scenario \cite{Randall:1999vf} which is based on only one brane (with the other one formally taken to be at infinity). The fundamental 6D braneworld model under our consideration may thus be interpreted as a combination of individual RS1 and RS2 scenarios along the two extra dimensions. Moreover, according to the analysis in \cite{Bhaumik:2023tmg}, this interpretation holds for any arbitrary $n$-fold nested warped geometry, where the model is analogous to a $(4+n)$-dimensional RS1 scenario only along the first orbifold, and mimics $(4+n)$-dimensional RS2 scenarios along all the subsequent (higher) orbifolds.

\section{Effective four-dimensional dynamics} \label{sec:modpot}
The 6D action in \eqref{eq:fullact}, when integrated over the compact spaces, results in a 4D effective action defined in terms of the two moduli $c$ and $k$. To address the full scope of the lower dimensional phenomenology resulting from such a higher dimensional framework, these moduli need to be treated as 4D scalar fields, \textit{\textit{i.e.}} $c(x)$ and $k(x)$. This is completely equivalent to working with $T_1(x)$ and $T_2(x)$, owing to the two constraint equalities relating the warping moduli and the orbifold radii. Consequently, $\mathcal{R}_6$ contains both the moduli fields and their 4D derivatives, with the former going into the effective potential and the latter defining the kinetic terms, once the extra dimensions have been integrated out. The rest of the potential is pieced together from the other terms of \eqref{eq:fullact}, which contain no derivatives of $c(x)$ and $k(x)$. In what follows, we focus on explicitly constructing the effective action by treating the potential and the kinetic terms separately.

Before proceeding directly with these calculations however, it is imperative to highlight how the effective 4D Planck scale ($M_{\rm eff}$) is connected to $M$, \textit{\textit{i.e.}} the fundamental Planck scale. This relation sets the strength of the couplings in the effective 4D theory and serves as a bridge between the latter and the fundamental higher-dimensional theory. As shown in \cite{Choudhury:2006nj}, the relation between $M_{\rm eff}$ and $M$ in the flat doubly warped braneworld model depends on the moduli $c$ and $k$, and can be expressed as
\begin{equation} \label{eq:planckrelflat}
    M_{\rm eff}^2=\dfrac{2M^4}{k'^2}\cosh(k\pi)\left(1-e^{-2c\pi}\right)\left(1+\dfrac{1}{3}\sinh^2(k\pi)\right)\tanh(k\pi)\sech(k\pi)\:,
\end{equation}
where the mass scale $k'=\sqrt{-\dfrac{\Lambda_6}{40M^4}}$ has been defined. With $\Lambda_6\sim-M^6$, we end up with $k'\sim M$. Thus, for dominant warping along either orbifold direction, \textit{\textit{i.e.}} with either $c\gtrsim\mathcal{O}(1)$ or $k\gtrsim\mathcal{O}(1)$, it is evident from \eqref{eq:planckrelflat} that the effective 4D Planck mass remains very close to the fundamental Planck mass $-$ a feature similar to the standard 5D scenario \cite{Randall:1999ee}. In the generalized non-flat scenario under our consideration, the 4D Planck mass can be obtained in a similar manner, being defined by
\begin{equation} \label{eq:planckreldef}
    M_{\rm eff}^2=M^4T_1T_2\int\limits_{-\pi}^{+\pi}dy\:a(y)^2\int\limits_{-\pi}^{+\pi}dz\:b(z)^3\:.
\end{equation}
which leads to different relations in the dS and AdS branches according to \eqref{eq:warpfacsads} and \eqref{eq:warpfacsds}. These relations can be analytically simplified to
\begin{itemize}
    \item AdS: 
    \begin{eqnarray}
     M_{\rm eff}^2=\dfrac{1}{12c_1^2}\left(\dfrac{M^4}{k'^2}\right)&&\times\left[\left(1-e^{-2c\pi}\right)c_1^4+4\pi c\:w_1^2c_1^2+\left(e^{2c\pi}-1\right)w_1^4\right] \nonumber \\
     &&\times\left[5+\cosh(2k\pi)\right]\tanh(k\pi)\sech(k\pi)\:,
    \end{eqnarray}
    \item dS:
    \begin{eqnarray}
    M_{\rm eff}^2=\dfrac{1}{12c_2^2}\left(\dfrac{M^4}{k'^2}\right)&&\times\left[\left(1-e^{-2c\pi}\right)c_2^4-4\pi c\:w_2^2c_2^2+\left(e^{2c\pi}-1\right)w_2^4\right] \nonumber \\
    &&\times\left[5+\cosh(2k\pi)\right]\tanh(k\pi)\sech(k\pi)\:.
    \end{eqnarray}
\end{itemize}
For $w_{1,2}\ll1$ and $c_{1,2}=1+\sqrt{1\mp w_{1,2}^2}$ as defined earlier, it is clear that dominant warping along either orbifold results in $M_{\rm eff}\approx M$, which is similar to the flat case. This proximity of the effective Planck scale to the fundamental scale is, indeed, a generic feature of warped braneworlds. Motivated by this, we do not distinguish explicitly between $M_{\rm eff}$ and $M$ in subsequent calculations.

\subsection{Origin of the induced moduli potential} \label{subsec:modpotorig}
Let us focus on the potential first. The extra-dimensional part of $\mathcal{R}_6$ which contributes to the 4D effective potential is given by 
\begin{equation}
\mathcal{R}_6^{(\textrm{pot})}=-\dfrac{8}{T_1^2b^2}\left(\dfrac{a''}{a}\right)-\dfrac{12}{T_1^2b^2}\left(\dfrac{a'}{a}\right)^2-\dfrac{10}{T_2^2}\left(\dfrac{b''}{b}\right)-\dfrac{20}{T_2^2}\left(\dfrac{b'}{b}\right)^2\:,
\end{equation}
where the primes denote derivatives with respect to (w.r.t.) the appropriate angular coordinates, \textit{\textit{i.e.}} w.r.t. $y$ for $a(y)$ and w.r.t. $z$ for $b(z)$. The dimensionless moduli potential $V(c,k)$ can then be defined as 
\begin{equation} \label{eq:spot}
    \begin{split}
        \mathcal{S}_{\textrm{eff}}^{(\textrm{pot})}\equiv-2M^4\int d^4x\sqrt{-g}V(c,k)=&\int d^4x\int\limits_{-\pi}^{+\pi}dy\int\limits_{-\pi}^{+\pi}dz\left[\sqrt{-g_6}\left(2M^4\mathcal{R}_6^{(\textrm{pot})}-\Lambda_6\right) \right. \\
        &-\sqrt{-g_5^{(\textrm{ind})}}\{V_1(z)\delta(y)+V_2(z)\delta(y-\pi)\} \\
        & \left. -\sqrt{-\bar{g}_5^{(\textrm{ind})}}\{V_3(y)\delta(z)+V_4(y)\delta(z-\pi)\}\right]\:.
    \end{split}
\end{equation} 
The relevant metric determinants are given explicitly by $\sqrt{-g_6}=T_1T_2a^4b^5\sqrt{-g}$, $\sqrt{-g_5^{(\textrm{ind})}}=T_2a^4b^4\sqrt{-g}$, and $\sqrt{-\bar{g}_5^{(\textrm{ind})}}=T_1a^4b^5\sqrt{-g}$. Substituting the explicit forms of the warp factors and the brane tensions, and integrating over $y$ and $z$, lead to the 4D effective potential for both the AdS and the dS case. For the explicit analytical form of $V(c,k)$, the interested reader is referred to Appendix \ref{sec:appendixA}. It is worth stressing again that a flat background with $\Omega=0$ makes this effective potential vanish identically, whereas one ends up with a non-zero contribution only for $\Omega\neq0$.

\subsection{Effective kinetic and curvature terms} \label{subsec:curvkin}
Apart from $\mathcal{R}_6^{\textrm{(\textrm{pot})}}$, the rest of $\mathcal{R}_6$ contains derivatives of the moduli fields with respect to the 4D brane coordinates $x^\mu$. These constitute $\mathcal{R}_6^{(\textrm{kin})}$, which is to be integrated to obtain the 4D effective kinetic terms for the moduli fields. To make the task at hand easier, it is worth focusing first on the $g_{\mu\nu}=\eta_{\mu\nu}$ case, for which this integrand is given by
\begin{equation}
\begin{split}
\mathcal{R}_6^{(\textrm{kin})}=&-\dfrac{6}{a^3b^2}\partial_\mu\partial^\mu a-\dfrac{8}{a^2b^3}\partial_\mu\partial^\mu b-\dfrac{2}{a^2b^2T_1}\partial_\mu\partial^\mu T_1-\dfrac{2}{a^2b^2T_2}\partial_\mu\partial^\mu T_2\\
& -\dfrac{4}{a^2b^4}\partial_\mu b\partial^\mu b-\dfrac{16}{a^3b^3}\partial_\mu a\partial^\mu b-\dfrac{4}{a^3b^2T_1}\partial_\mu T_1\partial^\mu a-\dfrac{4}{a^3b^2T_2}\partial_\mu T_2\partial^\mu a \\
& -\dfrac{8}{a^2b^3T_1}\partial_\mu T_1\partial^\mu b-\dfrac{6}{a^2b^3T_2}\partial_\mu T_2\partial^\mu b-\dfrac{2}{a^2b^2T_1T_2}\partial_\mu T_1\partial^\mu T_2\:.
\end{split}
\end{equation}
Taking into account $\sqrt{-g_6}=T_1T_2a^4b^5$ and integrating the double derivative terms by parts, the integral can be re-written as
\begin{equation} \label{eq:Rkin}
\begin{split}
\int d^4x\sqrt{-g_6}\mathcal{R}_6^{(\textrm{kin})} =&\int d^4x\sqrt{-g}\left[ 6b^3T_1T_2\partial _\mu a\partial ^\mu a \nonumber +18ab^2T_1T_2\partial _\mu a \partial ^\mu b \right. \\
&+12a^2bT_1T_2\partial _\mu b\partial ^\mu b \nonumber +6ab^3\partial _\mu a(T_2\partial ^\mu T_1+T_1\partial ^\mu T_2) \\
&\left. +2a^2b^2\partial _\mu b(3T_2\partial ^\mu T_1 \nonumber +4T_1\partial ^\mu T_2)+2a^2b^3\partial _\mu T_1\partial ^\mu T_2 \right]\:,
\end{split}
\end{equation}
where $\sqrt{-g}=1$ for $g_{\mu\nu}=\eta_{\mu\nu}$. On the contrary, if $g_{\mu\nu}\neq\eta_{\mu\nu}$, then $\sqrt{-g}$ remains explicitly in this expression as it is, and the rest of the contribution to $\mathcal{R}_6$, coming from the 4D curvature $R_{\mu\nu}$, splits off from both the kinetic and potential parts and appears as an additional term of form $g_6^{\mu\nu}R_{\mu\nu}=R/(a^2b^2)\in\mathcal{R}_6$. This can also be understood from the fact that the full 6D action must contain the lower dimensional Einstein-Hilbert action within itself. Hence, the additional integrand term turns out to be $\sqrt{-g_6}\times R/(a^2b^2)=T_1T_2a^2b^3\sqrt{-g}R$, which, when integrated over $y$ and $z$, gives the final piece in the 4D effective action as
\begin{equation} \label{eq:scurv}
\mathcal{S}_{\textrm{eff}}^{(\textrm{curv})}=2M^4\int d^4x T_1T_2\sqrt{-g}R\int\limits_{-\pi}^{+\pi}dy\int\limits_{-\pi}^{+\pi}dz a^2b^3=\dfrac{2M^4}{k'^2}\int d^4x\sqrt{-g}\:h(c,k)R\:,
\end{equation}
with the mass scale $k'=\sqrt{-\dfrac{\Lambda_6}{40M^4}}$ extracted to make the Jordan frame coupling $h(c,k)$ dimensionless. The explicit forms of $h(c,k)$ for the AdS and dS cases, obtained by plugging in the corresponding warp factors in the integral above, are given in \eqref{appeq:adsh} and \eqref{appeq:dsh} in Appendix \ref{sec:appendixA}.

With this clarification in place, one may now proceed with the simplification of the kinetic part from \eqref{eq:Rkin}. Plugging the warp factors for the non-flat cases into \eqref{eq:Rkin} and eliminating $T_1$ and $T_2$ algebraically in terms of $c$ and $k$, it becomes clear that the 4D effective kinetic action contains terms proportional to $\partial_\mu c \partial^\mu c$, $\partial_\mu k \partial^\mu k$, and $\partial_\mu c \partial^\mu k$. The action, however, is manifestly non-canonical, as the coefficients of these kinetic terms are functions of $c(x)$ and $k(x)$ that arise from integrating \eqref{eq:Rkin} appropriately over $y$ and $z$. Consequently, the kinetic action assumes the following form in the Jordan frame:
\begin{equation} \label{eq:skin}
    \begin{split}
        S_{\textrm{eff}}^{(\textrm{kin})}&=2M^4\int d^4x\int\limits_{-\pi}^{+\pi}dy\int\limits_{-\pi}^{+\pi}dz \sqrt{-g_6}\mathcal{R}_6^{(\textrm{kin})} \\
        &=\dfrac{2M^4}{k'^2}\int d^4x\sqrt{-g}\left(\tilde{A}_{cc}\partial_\mu c \partial^\mu c+\tilde{A}_{kk}\partial_\mu k \partial^\mu k+\tilde{A}_{ck}\partial_\mu c \partial^\mu k \right)\:,
    \end{split}
\end{equation}
where the explicit forms of the non-canonical coefficients $\tilde{A}_{cc}(c,k)$, $\tilde{A}_{kk}(c,k)$, and $\tilde{A}_{ck}(c,k)$ are highlighted in Appendix \ref{sec:appendixA}.

\subsection{Transformation from Jordan to Einstein frame} \label{subsec:jordeinst}
Stitching together the three distinct pieces from \eqref{eq:spot}, \eqref{eq:scurv}, and \eqref{eq:skin}, the total 4D effective action in the Jordan frame reads
\begin{equation}
    \begin{split}
        \mathcal{S}_{\textrm{eff}}^{\textrm{(tot)}}=&\:\:\dfrac{2M^4}{k'^2}\int d^4x\sqrt{-g}\:h(c,k)R-2M^4\int d^4x\sqrt{-g}\:V(c,k) \\
        &+\dfrac{2M^4}{k'^2}\int d^4x\sqrt{-g}\left(\tilde{A}_{cc}\partial_\mu c \partial^\mu c+\tilde{A}_{kk}\partial_\mu k \partial^\mu k+\tilde{A}_{ck}\partial_\mu c \partial^\mu k \right)\:.
    \end{split}
\end{equation}
Before proceeding further, some clarifications regarding the dimensions of the various quantities are in order. In the first integral, $[R]=+2$. In the second integral, the modular potential $V$ is dimensionless by construction. Finally, in the third integral, the non-canonical coefficients are dimensionless as well, rendering each individual kinetic term of dimension $+2$. With $k'$ being the derived mass scale, the entire 4D effective action is thus dimensionless as required.
Clearly, this action contains a non-minimal coupling of curvature to the moduli fields through $h(c,k)$. This coupling can be made minimal by transforming to the Einstein frame, which is enacted by first assuming 
\begin{equation}
    \hat{R}=\dfrac{R}{\psi^2}-\dfrac{6}{\psi^3}g^{\mu\nu}\nabla_\mu\nabla_\nu\psi\:,
\end{equation}
where $\hat{R}$ is the 4D Ricci scalar in the Einstein frame and $\psi(x)$ is an auxiliary field. The non-minimal coupling function can then be removed in the Einstein frame through the conformal transformation $\hat{g}_{\mu\nu}=\psi^2g_{\mu\nu}$, where for our purpose we make the choice $\psi(x)=\sqrt{h(c,k)}$. However, the same transformation also entails an additional contribution to the kinetic part as follows. Firstly, the curvature part of the action in the Einstein frame is calculated to be
\begin{equation}
    \mathcal{S}_{\textrm{eff}}^{(\textrm{curv})}=\dfrac{2M^4}{k'^2}\int d^4x\sqrt{-\hat{g}}\hat{R}+\dfrac{12M^4}{k'^2}\int d^4x\sqrt{-g}\psi\Box\psi\:,
\end{equation}
where $\Box\psi=g^{\mu\nu}\nabla_\mu\nabla_\nu\psi=\partial_\mu\partial^\mu\psi+\dfrac{1}{\sqrt{-g}}\partial_\mu(\sqrt{-g}\partial^\mu\psi)$. Then, integrating the second term of $\mathcal{S}_{\textrm{eff}}^{(\textrm{curv})}$ by parts, a kinetic contribution for $\psi$ emerges as follows:
\begin{equation}
    \dfrac{12M^4}{k'^2}\int d^4x\sqrt{-g}\psi\Box\psi\:\:\longrightarrow\:\:-\dfrac{12M^4}{k'^2}\int d^4x\sqrt{-g}\partial_\mu\psi\partial^\mu\psi\:.
\end{equation}
But since one has $\psi=\sqrt{h(c,k)}$, this term consists of quadratic terms in the derivatives of $c(x)$ and $k(x)$, resulting in a net kinetic expression that acts as a correction term to the original $\mathcal{S}_{\textrm{eff}}^{(\textrm{kin})}$ in the Einstein frame. The total Einstein frame action thus reduces to 
\begin{equation} \label{eq:seff}
    \begin{split}
        \mathcal{S}_{\textrm{eff}}^{(\textrm{tot})}=&\:\:\dfrac{2M^4}{k'^2}\int d^4x\sqrt{-\hat{g}}\:\hat{R}-2M^4\int d^4x\sqrt{-\hat{g}}\:\hat{V}(c,k) \\
        &+\dfrac{M^4}{k'^2}\int d^4x\sqrt{-g}\left(\hat{\tilde{A}}_{cc}\partial_\mu c \partial^\mu c+\hat{\tilde{A}}_{kk}\partial_\mu k \partial^\mu k+\hat{\tilde{A}}_{ck}\partial_\mu c \partial^\mu k \right)\:,
    \end{split}
\end{equation}
where $\hat{V}(c,k)=V(c,k)/h(c,k)^2$ is the potential in the Einstein frame, and the indices have been raised with respect to the Einstein frame metric $\hat{g}_{\mu\nu}$. The Ricci scalar has been rid of its direct coupling to the moduli fields. However, the kinetic terms still have non-minimal metric couplings, which are related to the old non-canonical coefficients by
\begin{equation} \label{eq:kinrels}
    \hat{\tilde{A}}_{cc}=\dfrac{2\tilde{A}_{cc}}{h}-\dfrac{3}{h^2}\left(\dfrac{\partial h}{\partial c}\right)^2\:,\:\:\hat{\tilde{A}}_{kk}=\dfrac{2\tilde{A}_{kk}}{h}-\dfrac{3}{h^2}\left(\dfrac{\partial h}{\partial k}\right)^2\:,\:\:\hat{\tilde{A}}_{ck}=\dfrac{2\tilde{A}_{cc}}{h}-\dfrac{6}{h^2}\left(\dfrac{\partial h}{\partial c}\right)\left(\dfrac{\partial h}{\partial k}\right)\:.
\end{equation}
The derivatives of $h(c,k)$ which appear in these relations encapsulate the aforementioned kinetic corrections arising out of the conformal transformation. 
\begin{figure}[!t]
    \centering
    \subfigure[]{\includegraphics[width=0.48\textwidth]{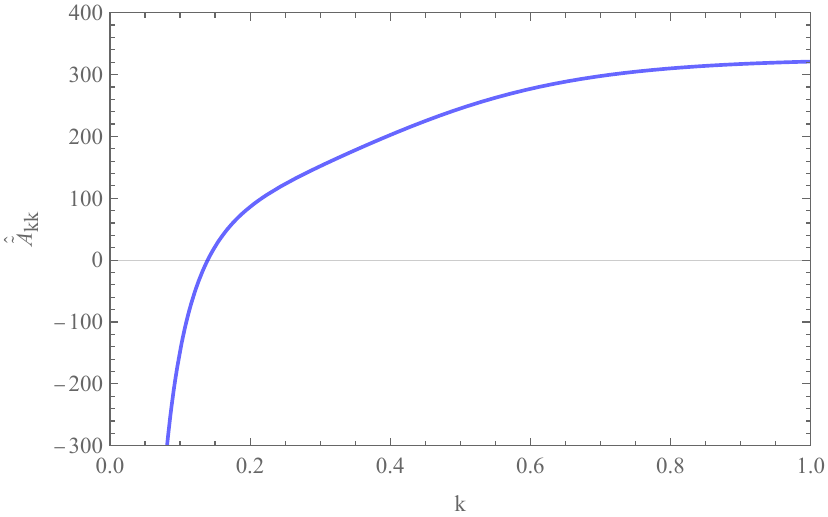}}
    \hfill
    \subfigure[]{\includegraphics[width=0.48\textwidth]{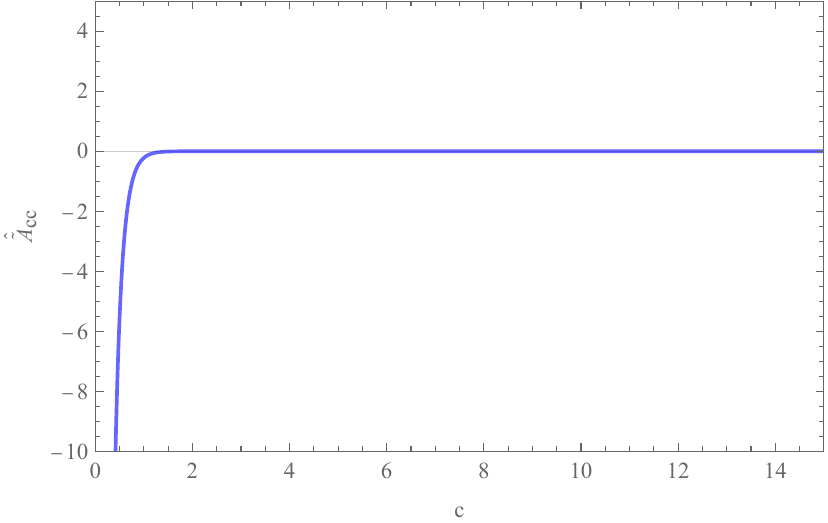}}
    \subfigure[]{\includegraphics[width=0.48\textwidth]{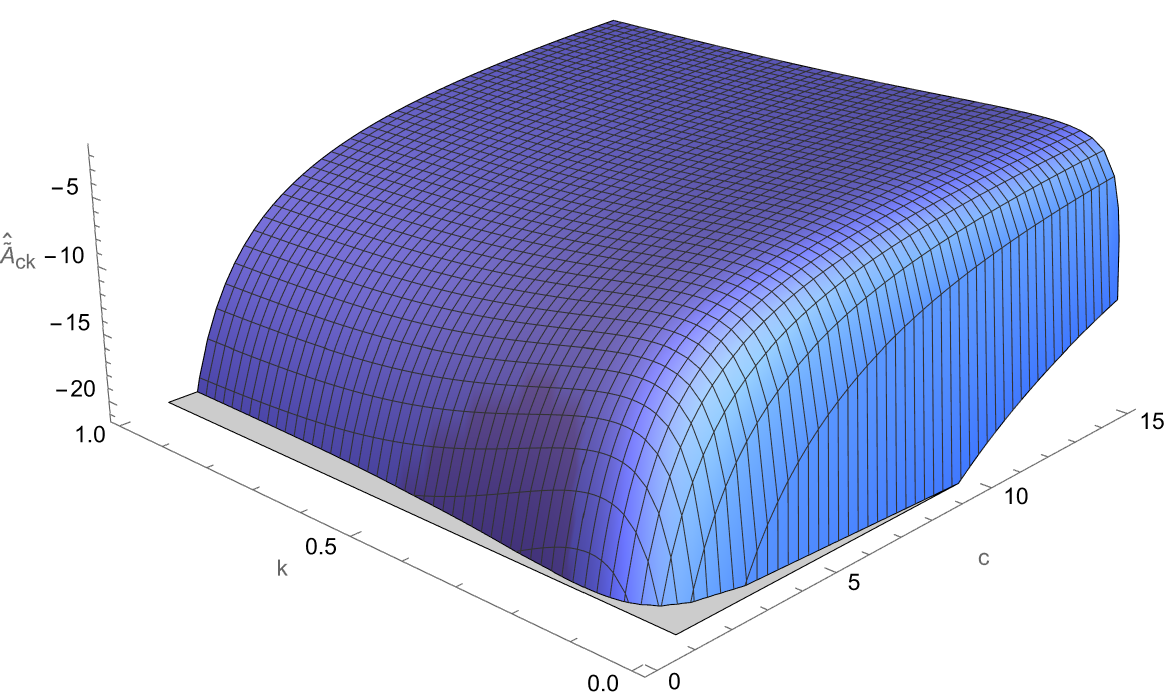}}
    \hfill
    \subfigure[]{\includegraphics[width=0.48\textwidth]{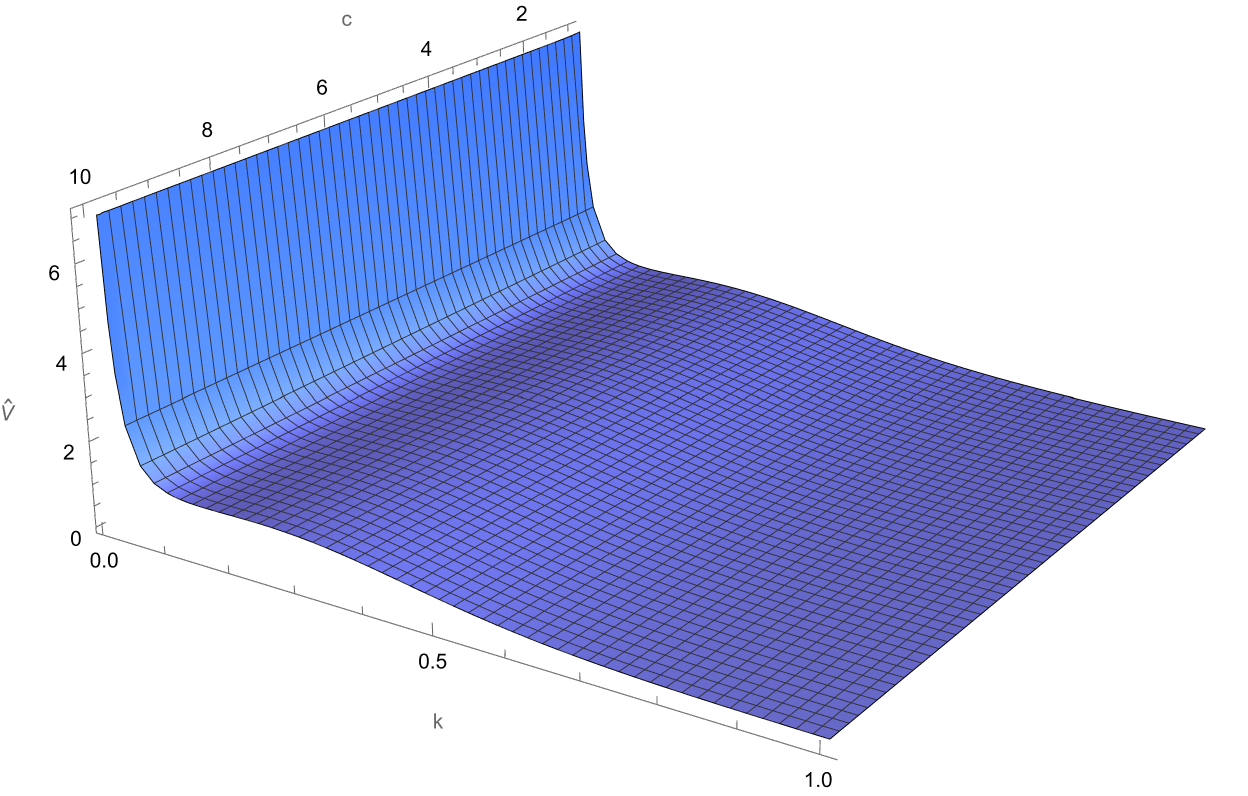}}
    \caption{Representative plots of the Einstein frame non-canonical kinetic coupling functions $\hat{\tilde{A}}_{kk}(k)$ on the upper left, $\hat{\tilde{A}}_{cc}(c)$ on the upper right, $\hat{\tilde{A}}_{ck}(c,k)$ at the lower left, and the Einstein frame induced potential $\hat{V}(c,k)$ at the lower right, for AdS geometry with $\omega_1=10^{-15}$. Note that all the quantities are dimensionless, and $\hat{V}(c,k)$ has been shown multiplied by an overall factor of $10^{15}$.}
    \label{fig:adsset}
\end{figure}
\begin{figure}[!t]
    \centering
    \subfigure[]{\includegraphics[width=0.48\textwidth]{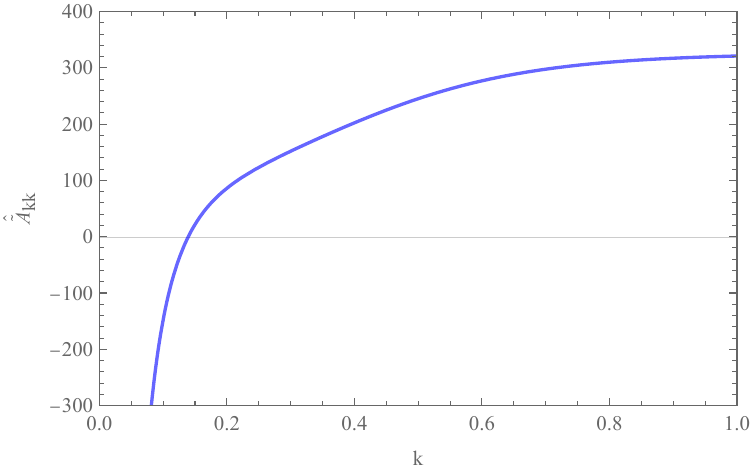} \label{subfig:dSAkk}}
    \hfill
    \subfigure[]{\includegraphics[width=0.48\textwidth]{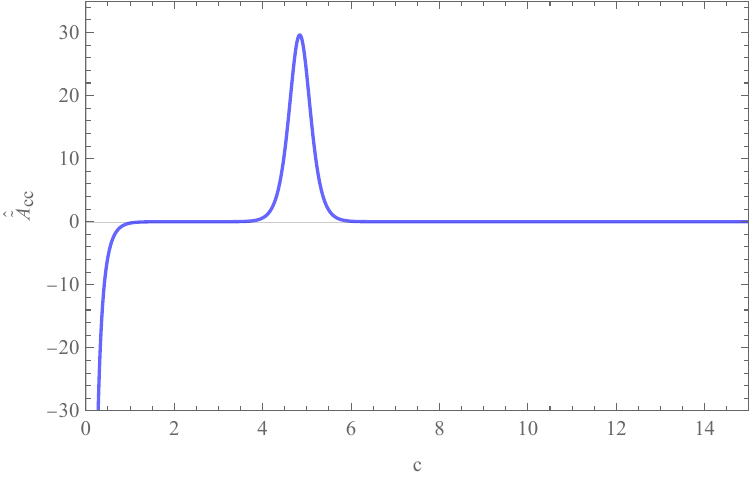} \label{subfig:dSAcc}}
    \subfigure[]{\includegraphics[width=0.48\textwidth]{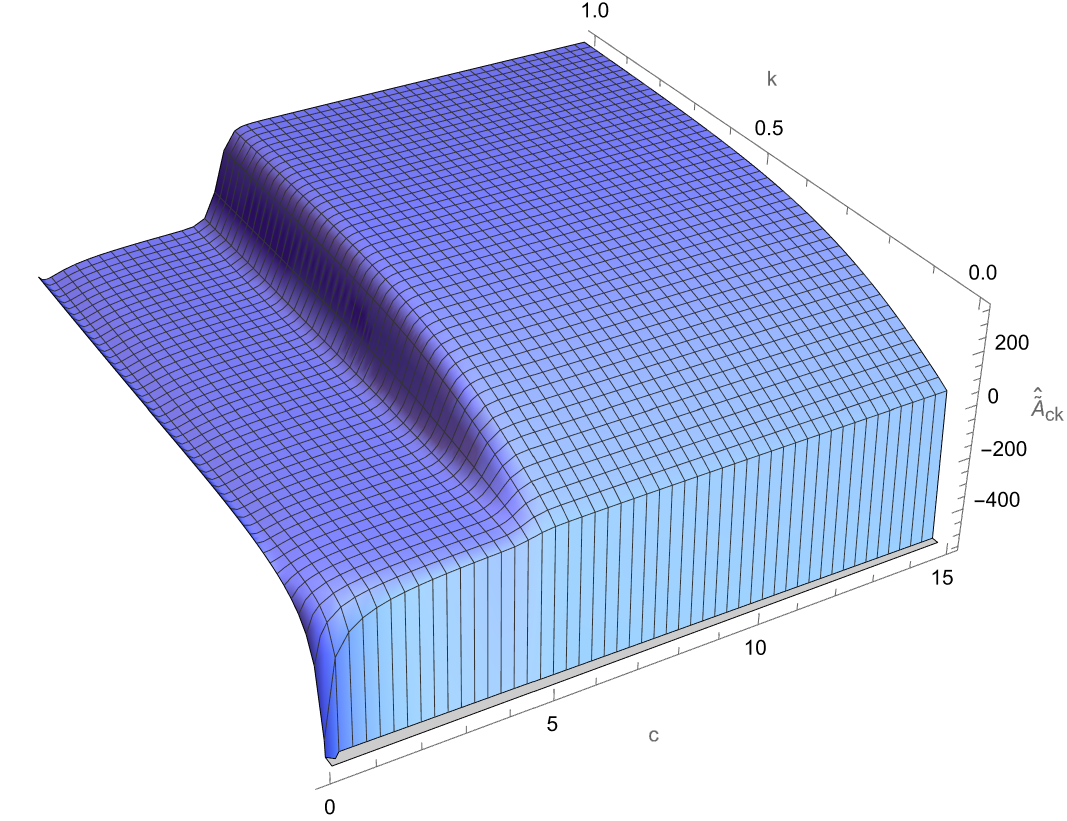} \label{subfig:dSAck}}
    \hfill
    \subfigure[]{\includegraphics[width=0.48\textwidth]{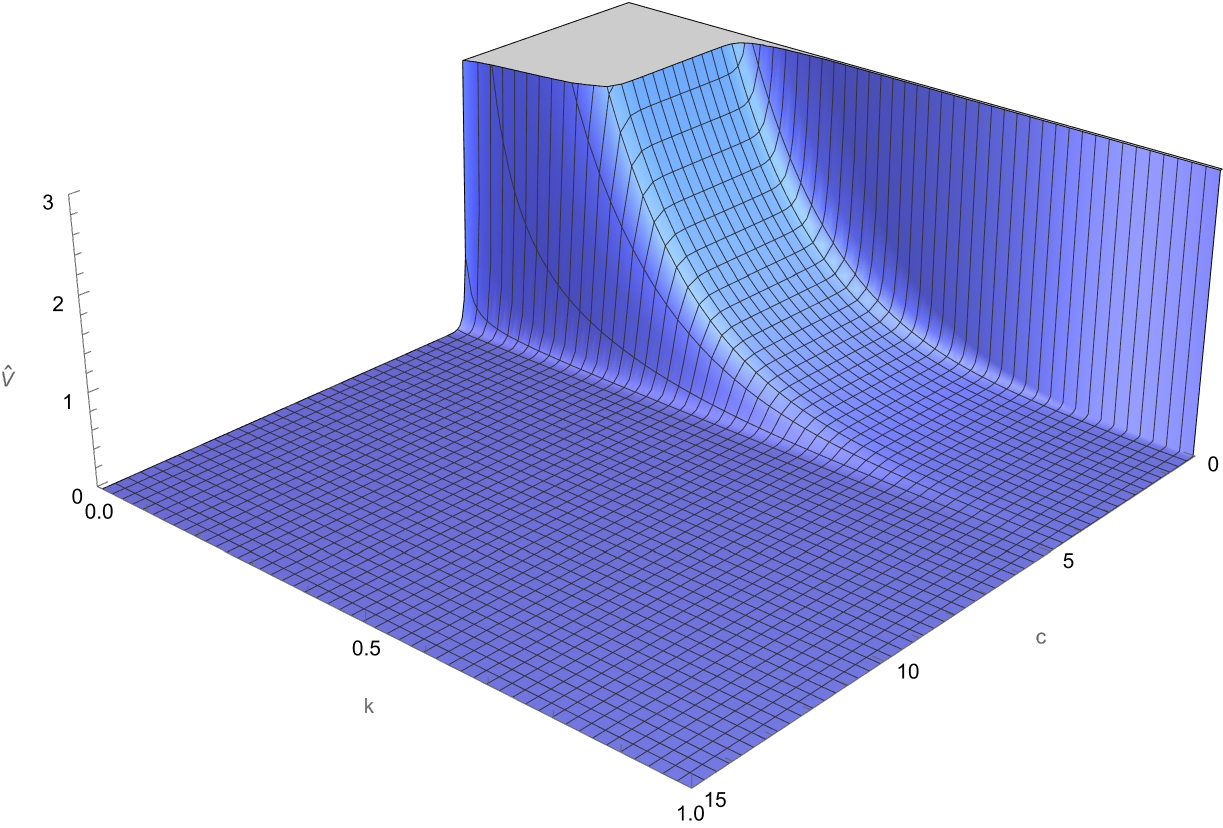} \label{subfig:dSV}}
    \caption{Representative plots of the Einstein frame non-canonical kinetic coupling functions $\hat{\tilde{A}}_{kk}(k)$ on the upper left, $\hat{\tilde{A}}_{cc}(c)$ on the upper right, $\hat{\tilde{A}}_{ck}(c,k)$ at the lower left, and the Einstein frame induced potential $\hat{V}(c,k)$ at the lower right, for dS geometry with $\omega_2=10^{-3}$. Note that all the quantities are dimensionless, and $\hat{V}(c,k)$ has been shown multiplied by an overall factor of $10^{6}$.}
    \label{fig:dsset}
\end{figure}
The kinetic coefficients and the induced potential in the Einstein frame have been plotted in Fig. \ref{fig:adsset} for the AdS case and in Fig. \ref{fig:dsset} for the dS case. The explicit forms of the AdS kinetic coefficients are given in Appendix \ref{sec:appendixA} under \eqref{appeq:adsacc}, \eqref{appeq:adsakk}, and \eqref{appeq:adsack}, and the dS ones under \eqref{appeq:dsacc}, \eqref{appeq:dsakk}, and \eqref{appeq:dsack}. The corresponding functional forms of $\hat{V}(c,k)$ appear in \eqref{appeq:adsmodpot} and \eqref{appeq:dsmodpot}. It is interesting to note that $\hat{\tilde{A}}_{cc}$ depends on $c$ only and $\hat{\tilde{A}}_{kk}$ is a function of $k$ alone, while $\hat{\tilde{A}}_{ck}$ is a function of both $c$ and $k$. In this sense, the Einstein frame allows a nice decomposition of the functional dependence of the coupling coefficients based on the kinetic terms they are associated with, which is a consequence of the underlying nested warped geometry and holds good in higher dimensional extensions. Also, $\hat{\tilde{A}}_{kk}(k)$ is independent of $\Omega$, and as such, is identical across the AdS and dS scenarios. These features appear to be generic properties which can be readily generalized to higher-dimensional extensions, as we discuss later in Sec. \ref{sec:gentrends}. On the other hand, $\hat{\tilde{A}}_{cc}(c)$ in the dS case changes sign from negative to positive (for increasing $c$) at a non-zero value of $c$, unlike in the AdS case where it approaches zero from the negative side asymptotically. This feature of the dS regime turns out to be crucial from a cosmological point of view, as we shall demonstrate in the upcoming section.

A few words about $\hat{V}(c,k)$ in the de Sitter regime are in order. In the 5D curved scenario, it is known that the induced dS potential may stabilize the extra-dimensional modulus by virtue of its metastable minimum \cite{Banerjee:2017jyk,Banerjee:2018kcz}. While these earlier studies have been carried out in terms of the canonically normalized radion field derived from the modulus, the metastable nature of the plateau should reflect in the modular potential as well. In fact, if one proceeds directly in terms of the modulus in the 5D set-up, the shape of the induced potential that one ends up with is identical to a constant $k$-projection of $\hat{V}(c,k)$ from Fig. \ref{subfig:dSV}, as expected physically. Such a constant $k$-slice clearly shows a nearly flat plateau in the range $c\in[1,5]$. For a given value of $\omega_2$, this range of $c$ where the metastable region occurs in the $c$-direction does not depend on the value of $k$, whose increment only results in an overall reduction in the height of the metastable interval as shown in Fig. \ref{subfig:dSV}. So here is the natural question to ask: might it be possible to stabilize the other moduli of multiply warped geometries in a similar fashion? Based on Fig. \ref{fig:dsset}, we find the answer to be negative. In fact, $\hat{V}(c,k)$ shows no local minimum, metastable or not, in the direction of the second modulus $k$, along which it decays monotonically. As we observe later in Sec. \ref{sec:gentrends}, this trend continues to hold in nested warped extensions beyond 6D, with the induced potential displaying no local minimum along the higher-dimensional moduli in field space. Thus, in multiply warped extensions of the 5D curved scenario, one indeed appears to need additional phenomenological mechanisms and/or beyond-Einstein gravitational effects to stabilize the higher moduli beyond the first one.

\subsection{Field space curvature in Einstein frame} \label{subsec:fieldcurv}

An important theoretical tool for illustrating the deviation from canonicity is the curvature of the so-called field space metric ($G_{AB}$), which consists of the non-canonical kinetic coefficients associated with the individual kinetic terms. Its definition allows expressing the overall kinetic part of the Lagrangian as $\mathcal{L}_{\rm kin}=G_{AB}\partial_\mu\phi^A\partial^\mu\phi^B$, where $A$ and $B$ label the individual fields. In our case, the full field space metric may be represented by a symmetric $2\times2$ matrix, which comprises of the following components:
\begin{eqnarray} \label{eq:fieldmetric}
    G_{11}=\hat{\tilde{A}}_{cc}(c)\:,\:\:G_{22}=\hat{\tilde{A}}_{kk}(k)\:,\:\:G_{12}=G_{21}=\dfrac{1}{2}\hat{\tilde{A}}_{ck}(c,k)\:,
\end{eqnarray}
where the first two terms are the diagonal terms associated respectively with $\partial_\mu c\partial^\mu c$ and $\partial_\mu k\partial^\mu k$, and the last term is the off-diagonal term associated with $\partial_\mu c\partial^\mu k$. Note that all the components are dimensionless.

\begin{figure}[!t]
    \centering
    \subfigure[]{\includegraphics[width=0.48\textwidth]{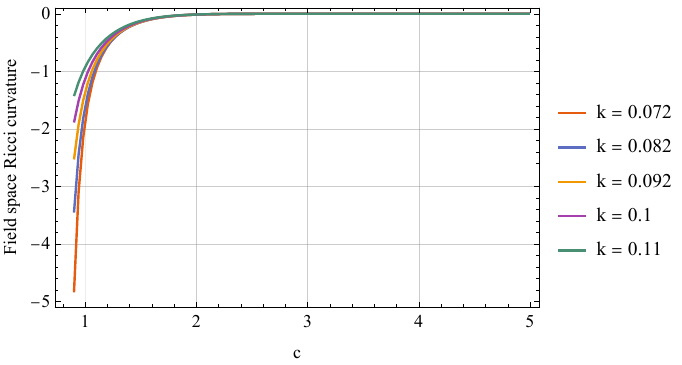}}
    \hfill
    \subfigure[]{\includegraphics[width=0.48\textwidth]{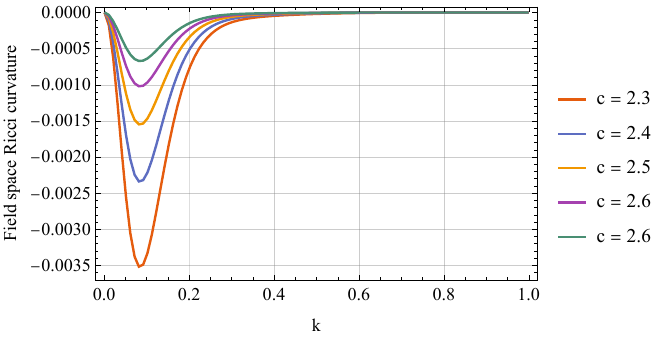}}
    \subfigure[]{\includegraphics[width=0.48\textwidth]{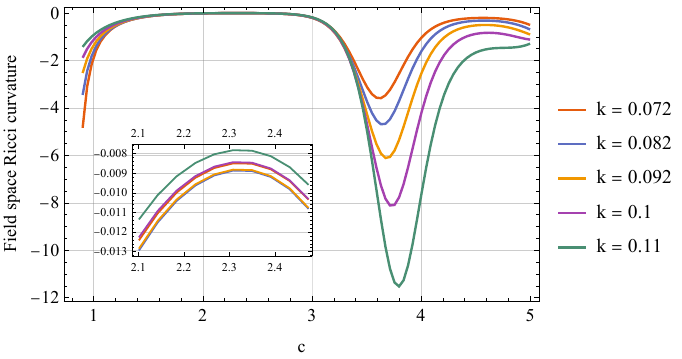}}
    \hfill
    \subfigure[]{\includegraphics[width=0.48\textwidth]{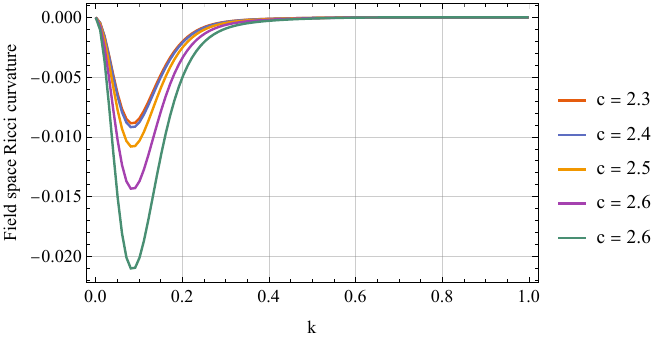}}
    \caption{Representative plots of the field space Ricci curvature scalar $\mathcal{R}_{\rm field}$ (dimensionless) in the AdS regime (\textit{upper row}) and in the dS regime (\textit{lower row}). Due to ease of visualization, one-dimensional profiles are shown as a function of one of the moduli fields at a time, with the variation with respect to the other one shown parametrically. Similar to Figs. \ref{fig:adsset} and \ref{fig:dsset}, the plots here correspond to $\omega_1=10^{-15}$ in the AdS case and $\omega_2=10^{-3}$ in the dS case.}
    \label{fig:fieldcurv}
\end{figure}

Using the tensor computation package xCoba (based on the Wolfram suite xAct \cite{xact}), the Ricci curvature scalar ($\mathcal{R}_{\rm field}$) associated with this $G_{AB}$, which is dimensionless owing to the dimensionless nature of $c$ and $k$, is obtained in an analytical form as
\begin{eqnarray} \label{eq:fieldcurv}
    \mathcal{R}_{\rm field}=&&\dfrac{4}{\left(\hat{\tilde{A}}_{ck}^2-4\hat{\tilde{A}}_{cc}\hat{\tilde{A}}_{kk}\right)^2}\Bigg[\hat{\tilde{A}}_{ck}\left(\dfrac{\partial\hat{\tilde{A}}_{cc}}{\partial c}\dfrac{\partial\hat{\tilde{A}}_{kk}}{\partial k}+\dfrac{\partial\hat{\tilde{A}}_{ck}}{\partial c}\dfrac{\partial\hat{\tilde{A}}_{ck}}{\partial k}\right) \nonumber \\
    &&-2\left(\hat{\tilde{A}}_{kk}\dfrac{\partial\hat{\tilde{A}}_{cc}}{\partial c}\dfrac{\partial\hat{\tilde{A}}_{ck}}{\partial k}+\hat{\tilde{A}}_{cc}\dfrac{\partial\hat{\tilde{A}}_{kk}}{\partial k}\dfrac{\partial\hat{\tilde{A}}_{ck}}{\partial c}\right)-\left(\hat{\tilde{A}}_{ck}^2-4\hat{\tilde{A}}_{cc}\hat{\tilde{A}}_{kk}\right)\dfrac{\partial^2 \hat{\tilde{A}}_{ck}}{\partial c\:\partial k}\Bigg]\:. \nonumber \\
    &&
\end{eqnarray}
We can use \eqref{eq:fieldcurv} to evaluate the field space curvature in both the AdS and the dS branches, by making use of the analytical expressions of the non-canonical kinetic coefficients obtained in each branch (\textit{viz.} Appendix \ref{sec:appendixA}). In Fig. \ref{fig:fieldcurv}, we present the behaviour of $\mathcal{R}_{\rm field}$ graphically. Owing to different order-of-magnitude variations in the directions of the two moduli fields, it is difficult to visualize the full $\mathcal{R}_{\rm field}$ surface as a function of both $c$ and $k$. Hence, we show one-dimensional profiles of $\mathcal{R}_{\rm field}$ as a function of one of the fields at a time, for a few benchmark parametric values of the other field. 

Some interesting features are readily observable in Fig. \ref{fig:fieldcurv}. Firstly, $\mathcal{R}_{\rm field}<0$ throughout the entire range of field values, which is a smoking gun signature of non-canonicity (canonical kinetic terms should yield $\mathcal{R}_{\rm field}=0$ identically). Secondly, in the AdS case, the scenario approaches canonicity monotonically as $c$ increases (\textit{\textit{i.e.}} $\mathcal{R}_{\rm field}\to0$), whereas the non-canonical nature becomes most pronounced for a finite value of $k$ (represented by the location of the trough in the upper-right plot) beyond which non-canonicity decreases monotonically along higher $k$. In the dS case, on the other hand, non-canonical behaviour becomes extremal for a combination of non-zero values of both $c$ and $k$, represented by the troughs along both the $c$ and the $k$ directions in the two plots in the bottom row of Fig. \ref{fig:fieldcurv}. Beyond these values, non-canonicity gradually decreases for higher values of $c$ and $k$, as the scenario asymptotes towards canonicity. Thirdly, while we have used the same range of parametric values for $c$ and $k$ in the AdS and dS cases, the variation of $\mathcal{R}_{\rm field}$ with respect to these values shows opposite trends to some extent across these two regimes. This becomes clear upon comparing the upper-right and the bottom-right plots of Fig. \ref{fig:fieldcurv}, where the variation of $\mathcal{R}_{\rm field}$ as a function of $k$ is shown for different parametric values of $c$. While the minima become shallower upon increasing the value of $c$ in the AdS case, they become deeper for the same in the dS case. Besides, the absolute value of $\mathcal{R}_{\rm field}$ at the minima are, on an average, one order of magnitude smaller in the AdS case than in the dS case for identical ranges of $c$.

\section{Prospects for inflation} \label{sec:inflation}
In this section, we analyze the role of the effective action from \eqref{eq:seff} in driving a period of non-canonical slow roll inflation which is consistent with current cosmological data. To that end, we shall focus exclusively on the dS regime throughout this section, as observations indicate that our very early Universe underwent a quasi-de Sitter phase of accelerated expansion. In order to proceed, we first need the 4D effective field equations for gravity (obtained by extremizing \eqref{eq:seff} w.r.t. $\hat{g}_{\mu\nu}$) and the equations of motion for the moduli fields (extremizing \eqref{eq:seff} w.r.t. $c$ and $k$). The former turns out to be
\begin{equation} \label{eq:efe}
    \begin{split}
        & \hat{R}_{\mu\nu}-\dfrac{1}{2}\hat{R}\hat{g}_{\mu\nu}+\dfrac{1}{2}\left[\hat{\tilde{A}}_{cc}\left(\partial_\mu c\partial_\nu c-\dfrac{1}{2}\hat{g}_{\mu\nu}\partial_\alpha c\partial^\alpha c\right)+\hat{\tilde{A}}_{kk}\left(\partial_\mu k\partial_\nu k-\dfrac{1}{2}\hat{g}_{\mu\nu}\partial_\alpha k\partial^\alpha k\right) \right. \\
        & \left. +\hat{\tilde{A}}_{ck}\left(\partial_{[\mu} c\partial_{\nu]} k-\dfrac{1}{2}\hat{g}_{\mu\nu}\partial_\alpha c\partial^\alpha k\right)\right] +\dfrac{k'^2}{2}\hat{g}_{\mu\nu}\hat{V}(c,k)=0\:,
    \end{split}
\end{equation}
where the $\partial_\mu c\partial_\nu k$ cross-term has been symmetrized to preserve the symmetry of the field equations under the interchange of $\mu$ and $\nu$, \textit{\textit{i.e.}} the energy-momentum tensor $\hat{T}_{\mu\nu}(c,k)$ associated with \eqref{eq:seff} has been made symmetric. Similarly, the equations of motion for $c(x)$ and $k(x)$ are given by
\begin{equation} \label{eq:eqmotc}
    2\hat{\tilde{A}}_{cc}\nabla_\mu\nabla^\mu c+\hat{\tilde{A}}_{ck}\nabla_\mu\nabla^\mu k+{\hat{\tilde{A}}_{cc}}_{,c}\:\partial_\mu c\partial^\mu c+{\hat{\tilde{A}}_{ck}}_{,k}\:\partial_\mu k\partial^\mu k+2k'^2\hat{V}_{,c}=0\:,
\end{equation}
\begin{equation} \label{eq:eqmotk}
    2\hat{\tilde{A}}_{kk}\nabla_\mu\nabla^\mu k+\hat{\tilde{A}}_{ck}\nabla_\mu\nabla^\mu c+{\hat{\tilde{A}}_{kk}}_{,k}\:\partial_\mu k\partial^\mu k+{\hat{\tilde{A}}_{ck}}_{,c}\:\partial_\mu c\partial^\mu c+2k'^2\hat{V}_{,k}=0\:,
\end{equation}
where the commas in the subscripts indicate partial derivatives. As a remark, \eqref{eq:eqmotc} and \eqref{eq:eqmotk} when written in this form appear symmetric under the symbolic interchange of $c$ and $k$, which is a nice property of the underlying geometry. However, this symmetry is only superficial as neither $\hat{\tilde{A}}_{ck}(c,k)$ nor $\hat{V}(c,k)$ is actually symmetric in their arguments $c$ and $k$. To make contact with cosmology, we now take $\hat{g}_{\mu\nu}$ to be the maximally symmetric Friedmann-Lema\^{i}tre-Robertson-Walker (FRLW) metric given by $ds^2=-dt^2+a(t)^2d\vec{x}_3^2$, where we have assumed spatial flatness. Homogeneity and isotropy render the background moduli fields only time-dependent. The field equations \eqref{eq:efe} then split off into the following pair of Friedmann equations:
\begin{equation} \label{eq:fried1}
    H^2=-\dfrac{1}{12}\left(\hat{\tilde{A}}_{cc}\dot{c}^2+\hat{\tilde{A}}_{kk}\dot{k}^2+\hat{\tilde{A}}_{ck}\dot{c}\dot{k}\right)+\dfrac{k'^2}{6}\hat{V}(c,k)\:,
\end{equation}
\begin{equation} \label{eq:fried2}
    \dot{H}=\dfrac{1}{4}\left(\hat{\tilde{A}}_{cc}\dot{c}^2+\hat{\tilde{A}}_{kk}\dot{k}^2+\hat{\tilde{A}}_{ck}\dot{c}\dot{k}\right)\:,
\end{equation}
whereas \eqref{eq:eqmotc} and \eqref{eq:eqmotk} reduce to
\begin{equation} \label{eq:friedmot1}
    2\hat{\tilde{A}}_{cc}\ddot{c}+\hat{\tilde{A}}_{ck}\ddot{k}+3H\left(2\hat{\tilde{A}}_{cc}\dot{c}+\hat{\tilde{A}}_{ck}\dot{k}\right)+{\hat{\tilde{A}}_{cc}}_{,c}\dot{c}^2+{\hat{\tilde{A}}_{ck}}_{,k}\dot{k}^2-2k'^2{\hat{V}}_{,c}=0\:,
\end{equation}
\begin{equation} \label{eq:friedmot2}
    2\hat{\tilde{A}}_{kk}\ddot{k}+\hat{\tilde{A}}_{ck}\ddot{c}+3H\left(2\hat{\tilde{A}}_{kk}\dot{k}+\hat{\tilde{A}}_{ck}\dot{c}\right)+{\hat{\tilde{A}}_{kk}}_{,k}\dot{k}^2+{\hat{\tilde{A}}_{ck}}_{,c}\dot{c}^2-2k'^2{\hat{V}}_{,k}=0\:,
\end{equation}
where $H=\dot{a}/a$ is the Hubble parameter, and the dots indicates time derivatives. 

While it may be possible to solve the full system \eqref{eq:fried1}$-$\eqref{eq:friedmot2} numerically with appropriate initial conditions, we make here a simplifying assumption that would allow us to focus on the essential features of the multiply warped geometry at the cost of a reduced number of degrees of freedom. We assume one of the moduli to be chiefly responsible for driving inflation, with negligible rolling in the direction of the other modulus. In effect, this amounts to treating the latter modulus as a fixed parameter throughout an effective single-field inflationary epoch, which may be the result of a suitably implemented stabilization scheme. While this scenario may, at first glance, appear to be a little \textit{ad hoc}, we remind the reader that the chief aim of the current work is not to propose a new model of non-canonical multi-field inflation \textit{per se}, but rather to make a particular case study of one possible cosmological facet of curved multiply warped geometries. That being said, even within the present framework alone, it is possible to physically motivate the assumed unidirectional nature of the slow-roll based on certain quantitative arguments which we discuss in Appendix \ref{sec:appendixC}. With this rationale in mind, we now proceed, with the aforementioned approximation, on to a case-by-case study of what each scenario implies for inflation and its observable features. 

\subsection{Inflation driven by $c$} \label{subsec:cinf}
In this domain, the derivatives of $k$ drop out of \eqref{eq:fried1}$-$\eqref{eq:friedmot1}, and \eqref{eq:friedmot2} becomes obsolete as the variation of \eqref{eq:seff} w.r.t. $k$ is no longer meaningful. Thus, $k$ appears only parametrically in $\hat{V}(c,k)$, while inflation results from the slow roll of $c(t)$ at a fixed $k$-slice. The role of $k$ is subtle but important, as it essentially acts as a regulator for the height of the inflationary potential, which is interesting from both theoretical and observational perspectives as we shall demonstrate shortly. The system of equations can be made dimensionless by changing the independent variable to the number of $e$-folds ($N$) counted from the beginning of inflation, defined as $dN=Hdt$. The derivatives transform accordingly as $\dot{c}=H\underaccent{\boldsymbol{\cdot}}{c}$ and $\ddot{c}=H\underaccent{\boldsymbol{\cdot}}{H}\underaccent{\boldsymbol{\cdot}}{c}+H^2\underaccent{\boldsymbol{\cdot}\boldsymbol{\cdot}}{c}$, where the underdot denotes derivative w.r.t. $N$. Defining the dimensionless Hubble parameter $\bar{H}=H/k'$ and using the Friedmann equations, the exact evolution of $c(N)$ is given by the second-order nonlinear ordinary differential equation (ODE)
\begin{equation} \label{eq:ceqexact}
    2\hat{\tilde{A}}_{cc}\left(\underaccent{\boldsymbol{\cdot}\boldsymbol{\cdot}}{c}+3\underaccent{\boldsymbol{\cdot}}{c}+\dfrac{1}{4}\hat{\tilde{A}}_{cc}\underaccent{\boldsymbol{\cdot}}{c}^3\right)+\hat{\tilde{A}}_{cc,c}\underaccent{\boldsymbol{\cdot}}{c}^2=12\left(\dfrac{\hat{V}_{,c}}{\hat{V}}\right)\left(1+\dfrac{1}{12}\hat{\tilde{A}}_{cc}\underaccent{\boldsymbol{\cdot}}{c}^2\right)\:,
\end{equation}
where we have not made any \textit{a priori} slow roll assumption. Solving this equation numerically requires knowledge of both $c(0)$ and $\underaccent{\boldsymbol{\cdot}}{c}(0)$, where for the latter we may assume slow roll to be valid at $N=0$ and obtain the initial rolling ``speed'' in terms of the initial field value ${c}(0)$. The first Hubble slow roll parameter $\epsilon(N)$ is given by
\begin{equation} \label{eq:epsc}
    \epsilon(N)\equiv-\dfrac{\dot{H}}{H^2}=-\dfrac{1}{4}\hat{\tilde{A}}_{cc}\underaccent{\boldsymbol{\cdot}}{c}^2\:\:\xrightarrow[]{\text{slow roll}}\:\:-\dfrac{1}{\hat{\tilde{A}}_{cc}}\left(\dfrac{\hat{V}_{,c}}{\hat{V}}\right)^2,
\end{equation}
with $\epsilon(N)\sim\mathcal{O}(1)$ marking the end of inflation. To address the horizon and flatness problems successfully, one typically needs the total number of $e$-folds to be $N_{\textrm{tot}}\gtrsim50-60$. 
\begin{figure}[!t]
    \centering
    \subfigure[]{\includegraphics[width=0.48\textwidth]{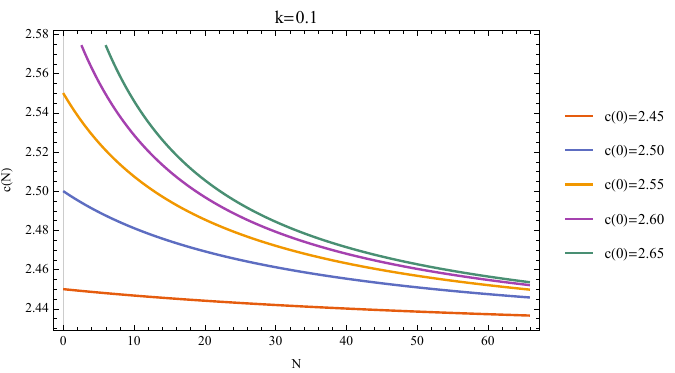}}
    \subfigure[]{\includegraphics[width=0.48\textwidth]{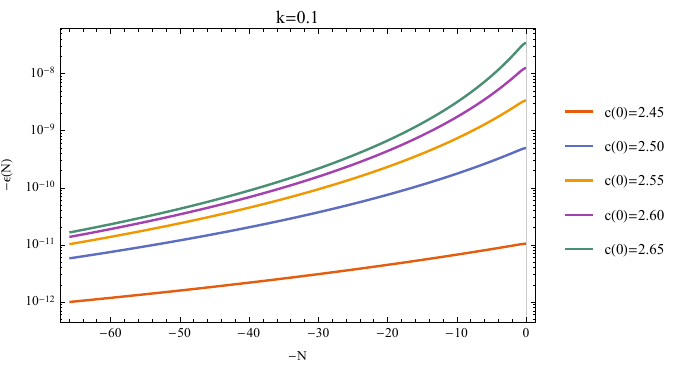}}
    \subfigure[]{\includegraphics[width=0.48\textwidth]{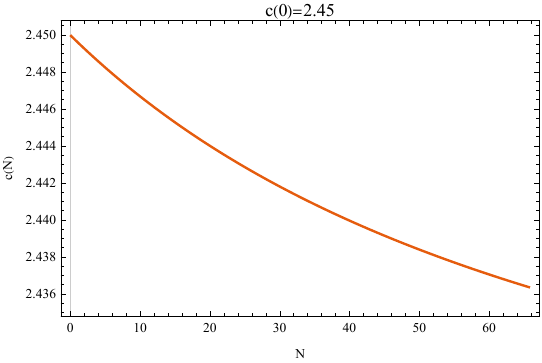}}
    \subfigure[]{\includegraphics[width=0.48\textwidth]{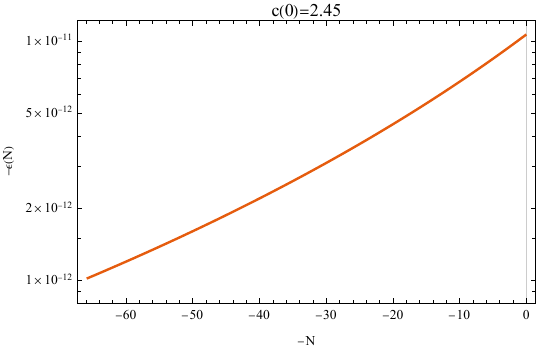}}
    \caption{Exact slow roll evolution of $c(N)$ and $\epsilon(N)$ plotted for $k=0.1$ and a few initial conditions $c(0)>c_{\rm in}\approx2.41994$, with the latter corresponding to $\omega_2=10^{-3}$ (upper panel). For a given $c(0)$, the slow roll trajectories are independent of the value of $k$ and coincide with each other (lower panel). This scenario is not of observational interest, as inflation does not come to an end at any finite $N_{\rm tot}$.}
    \label{fig:c1c2_exact}
\end{figure}
\begin{figure}[!t]
    \centering
    \subfigure[]{\includegraphics[width=0.48\textwidth]{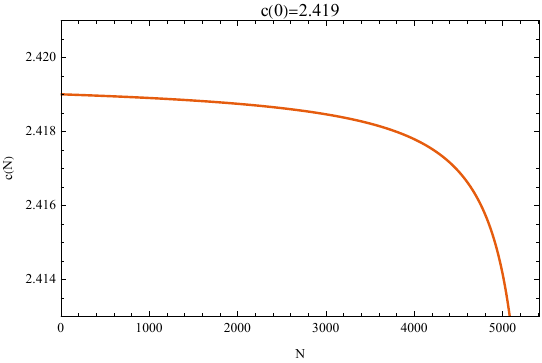}}
    \subfigure[]{\includegraphics[width=0.48\textwidth]{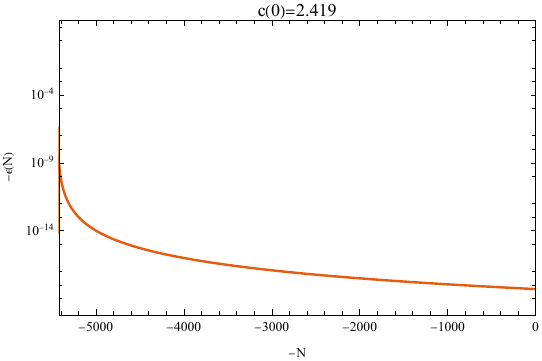}}
     \subfigure[]{\includegraphics[width=0.48\textwidth]{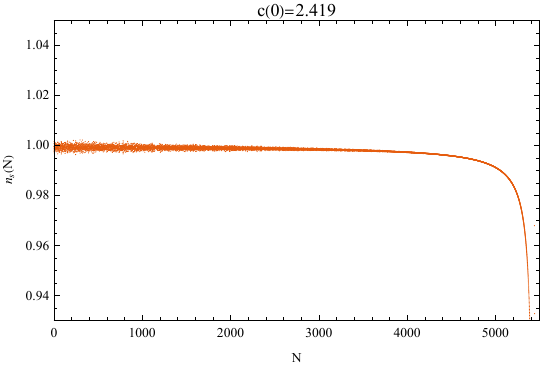}}
    \caption{Exact slow roll evolution of $c(N)$, $\epsilon(N)$, and $n_s(N)$ have been plotted for the initial condition $c(0)=2.419$, which results in $N_\textrm{tot}\approx5420$ for $\omega_2=10^{-3}$. The evolution is independent of the value of the parameter $k$, \textit{i.e.} trajectories for different plausible $k$-values (ranging between $10^{-3}$ and $10$) coincide. Note that in the upper-right plot, inflation proceeds from right to left, and the steep upward curve towards the left margin actually represents the divergence of $|\epsilon(N)|$ close to that point signifying the the end of inflation. While an extended vertical plot range better illustrates this divergence, it visually flattens the rest of the curve and makes it difficult to show the precise evolution of $\epsilon(N)$ throughout the course of inflation. The smearing of $n_s(N)$ in the bottom plot at smaller $e$-folds is a numerical artifact that gets resolved towards the tail end, which is of observational interest. This set of figures represents a viable scenario in the light of current CMB observations.}
    \label{fig:c3_exact}
\end{figure}

Using \eqref{eq:ceqexact}, we now focus on the direct computation of the field trajectories, for which the following observation proves useful first. At a constant $k$-slice, the de Sitter potential $\hat{V}(c,k)$ has a point of inflection along $c$ at $c_{\textrm{in}}=\ln(c_2/\omega_2)/\pi$, which, for $\omega_2=10^{-3}$, turns out to be $c_\textrm{in}\approx2.41994$. If the field starts close to this inflection point, one may have successful slow roll resulting in inflation. Now, the only input of $k$ which seemingly enters \eqref{eq:ceqexact} is through the term $(\hat{V}_{,c}/\hat{V})$. A numerical calculation of this ratio for a range of $k$-values, ranging from $10^{-3}$ to $10$, brings out a rather surprising result: $(\hat{V}_{,c}/\hat{V})$ is independent of $k$! In other words, the evolution of $c(N)$ is not parametrically dependent on the value of $k$, whose only role comprises of regulating the height of the potential plateau and its slope simultaneously in such a way that their ratio remains invariant. The nature of $(\hat{V}_{,c}/\hat{V})$ is shown in Fig. \ref{fig:app1}.

The dynamics differ markedly depending on whether the field starts on the left or right of the inflection point. For $c(0)>c_\textrm{in}$, inflation does not come to an end through the violation of slow roll at any finite $N_\textrm{tot}$. This is because the non-canonical kinetic coefficient $\hat{\tilde{A}}_{cc}$, which results in a phantom-like kinetic term in this domain, drives the field up the potential slope towards the inflection point, with a gradually decreasing rolling speed and a subsequently decreasing value of $\epsilon(N)$. This behaviour of $\epsilon(N)$ can be understood from \eqref{eq:epsc}, where $\epsilon(N)$ is proportional to $\hat{V}_{,c}$ under the slow roll assumption. Since the latter approaches zero (by definition) as $c\to c_\textrm{in}$, $\epsilon(N)$ does not diverge and inflation can continue \textit{ad infinitum}, unless some additional exit mechanism is invoked. This is displayed in Fig. \ref{fig:c1c2_exact} for a few benchmark values. On the other hand, for $c(0)<c_\textrm{in}$, inflation can come to an end naturally via slow roll violation. In this case, the phantom kinetic term makes $c(N)$ decrease over time in a similar manner as before, as the field is slowly driven up the potential plateau. However, $\hat{\tilde{A}}_{cc}$ changes its sign from positive to negative (for decreasing $c$) at a non-zero value of $c$ between $1$ and $c_\textrm{in}$, \textit{viz.} upper right panel of Fig. \ref{fig:dsset}. For $\omega_2=10^{-3}$, this critical value lies at $c\approx2.1$, where $\hat{\tilde{A}}_{cc}\to0$. As $\epsilon(N)$ is inversely proportional to $\hat{\tilde{A}}_{cc}$, this serves as the point where $|\epsilon(N)|$ diverges and inflation consequently ends, as shown in Fig. \ref{fig:c3_exact} for a case of physical interest. The upper-right plot of Fig. \ref{fig:c3_exact}, however, does not visually reflect this divergence to a high level of accuracy, and shows only a steep upward curve close to the end (left margin) instead. This upward curve actually represents the aforesaid divergence, which is hard to depict graphically in a clearer manner. While extending the vertical range of the plot can better show the divergence, extending it to $\mathcal{O}(1)$ (as required to concretely show the end of inflation) visually flattens the rest of the curve and makes the plot lose information about precisely how $\epsilon(N)$ varies throughout inflation. Hence, we present the $N$ versus $\epsilon(N)$ plot in the form that appears in Fig. \ref{fig:c3_exact}, keeping in mind that a divergence of the slow-roll parameter actually ends the inflating scenario as indicated by the steep upward rise of $|\epsilon(N)|$ close to $N_{\rm tot}\sim5420$. Just to clarify at this point once more, recall that the range of $e$-folds of observable interest in cosmology consists almost entirely of the final $50-60$ $e$-foldings before the end of inflation, which, in case of Fig. \ref{fig:c3_exact}, is comprised of $N\sim5360-5370$. However, we have plotted the evolution of $c(N)$, $\epsilon(N)$, and $n_s(N)$ throughout the entire course of inflation (which is markedly larger in our case than the physically observable range), in order to clearly illustrate the theoretical nature of the trajectories as well as to make it easier to connect Fig. \ref{fig:c3_exact} with the initial values chosen in our calculations. Moreover, as this scenario involves momentary vanishing of the kinetic term, the analytic prescription adopted here may not be completely trustworthy close to this endpoint. To account for this, we conservatively allow an error margin of $\sim0.2\%$ centered at the value of $N_\textrm{tot}$ where slow roll violation is naively expected. This poses little threat to our framework, as the total duration of inflation remains considerably larger than the minimum required amount. 

With the slow roll trajectories at hand, the scalar spectral index ($n_s$) can be estimated as a function of $e$-foldings up to the first order in slow roll as \cite{Garriga:1999hf,Chung:2003iu}
\begin{equation} \label{eq:nsc}
    n_s(N)\:\approx\:1-3\left(1+\dfrac{p_c}{\rho_c}\right)-\dfrac{1}{H}\dfrac{d}{dt}\left[\textrm{ln}\left(1+\dfrac{p_c}{\rho_c}\right)\right]=1+\dfrac{1}{2}\hat{\tilde{A}}_{cc}\underaccent{\boldsymbol{\cdot}}{c}^2-\left(\dfrac{\hat{\tilde{A}}_{cc,c}}{\hat{\tilde{A}}_{cc}}\right)\underaccent{\boldsymbol{\cdot}}{c}-2\left(\dfrac{\underaccent{\boldsymbol{\cdot}\boldsymbol{\cdot}}{c}}{\underaccent{\boldsymbol{\cdot}}{c}}\right)\:,
\end{equation}
where $p_c=\hat{X}-\hat{V}$ and $\rho_c=\hat{X}+\hat{V}$ are the pressure and energy densities of the $c$-field respectively (with $\hat{X}$ being the non-canonical kinetic term). To make contact with observables, $n_s$ is to be evaluated at the horizon-crossing of the mode of interest ($k_*$), which is defined by $k_*=a_*H_*$. For the case of the nearly constant potential $V_*\sim V_\textrm{end}\sim k'^2M^2\hat{V}$ throughout $c$-driven inflation and assuming prompt reheating, the horizon-crossing $e$-fold of $k_*\sim0.05$ Mpc\textsuperscript{-1} counted from the beginning of inflation is calculated to be \cite{Liddle:1993fq,Liddle:2003as,Leach:2002dw}
\begin{equation} \label{eq:efoldpart}
    N_{k_*}\approx N_{\textrm{tot}}-\left(66+0.25\ln\hat{V}\right)\:,
\end{equation}
where $k'\sim M$ has been assumed. In other words, the mode relevant to the cosmic microwave background (CMB) exits the horizon around $60$ $e$-folds before the end of inflation, which is compatible with well-accepted bounds. 

In Fig. \ref{fig:c3_exact}, we show the slow roll evolution for the initial condition $c(0)=2.419$, which results in $N_{\rm tot}\approx5420$ independently of the value of $k$. On the other hand, although $N_{k_*}$ is a function of $k$ via $\hat{V}$ according to \eqref{eq:efoldpart}, this dependence is logarithmic and hence extremely suppressed. In fact, for $0.001\lesssim k\lesssim 2.0$, the variation of $N_{k_*}$ in confined roughly to the range $5355\lesssim N_{k_*}\lesssim 5361$. While $N_{k_*}$ can grow faster for $k\gtrsim2.0$ due to a strong suppression of the magnitude of $\hat{V}$, such high $k$-values also imply a larger hierarchy $(T_1/T_2)_*\gtrsim\mathcal{O}(10^4)$ between the extra-dimensional radii at horizon-crossing, with the same being true for $k\lesssim 10^{-4}$. This is why we conservatively limit $k$ to the aforementioned range. To quote the values at these two extremities, we obtain $n_s(k_*)\approx0.9535$ for $k=2.0$, whereas we have $n_s(k_*)\approx0.9564$ for $k=0.001$. The former value falls close to the lower boundary of the $95\%$ confidence level (CL) interval of $n_s$ obtained from the CMB TT+TE+EE+lowE dataset of Planck 2018, whereas the latter falls within $95\%$ CL of the TT+TE+EE+lowE+lensing bounds \cite{Planck:2018nkj,Planck:2018vyg,Planck:2018jri}. In Table \ref{tab:table1}, we list these values alongside a few intermediate ones for appropriate combinations of the braneworld parameters. 
\begin{table}[!t]
    \centering
    \begin{tabular}{|c|c|c|c|c|c|c|}
    \hline
    $c(0)$ & $k$ & $(T_1/T_2)_*$ & $\hat{V}$ & $N_{k_*}$ & $n_s(k_*)$ \\
    \hline\hline
    $2.419$ & $0.001$ & $2394.34$ & $9.55\times10^{-4}$ & $5355.55$ & $0.9564$ \\ 
    \hdashline[2pt/3pt]
    $2.419$ & $0.01$ & $239.52$ & $9.54\times10^{-5}$ & $5356.33$ & $0.9560$ \\  
    \hdashline[2pt/3pt]
    $2.419$ & $0.1$ & $25.13$ & $9.09\times10^{-6}$ & $5357.01$ & $0.9556$ \\ 
    \hdashline[2pt/3pt]
    $2.419$ & $0.5$ & $12.01$ & $4.71\times10^{-7}$ & $5357.69$ & $0.9554$ \\ 
    \hdashline[2pt/3pt]
    $2.419$ & $0.9$ & $22.55$ & $1.45\times10^{-8}$ & $5358.62$ & $0.9548$ \\ 
    \hdashline[2pt/3pt]
    $2.419$ & $1.5$ & $88.80$ & $5.22\times10^{-11}$ & $5359.88$ & $0.9540$ \\ 
    \hdashline[2pt/3pt]
    $2.419$ & $2.0$ & $320.32$ & $4.69\times10^{-13}$ & $5361.04$ & $0.9535$ \\
    \hline
    \end{tabular}
    \caption{For the CMB pivot scale $k_*=0.05$ Mpc\textsuperscript{-1}, the interrelation among the inflationary quantities, \textit{i.e.} the CMB exit $e$-fold $N_{k_*}$ from the beginning of inflation and the scalar spectral index $n_s(k_*)$, and the braneworld parameters, \textit{i.e.} the initial value $c(0)$, the parameter $k$, and the extra-dimensional radii ratio at horizon-crossing $(T_1/T_2)_*$, has been shown for $c$-driven slow roll inflation. For values of $k$ outside this tabulated range, $(T_1/T_2)_*$ quickly grows larger than $\mathcal{O}(10^3)$, which amounts to a problematic hierarchy between the sizes of the extra dimensions. Recall that $\hat{V}(c,k)$ is dimensionless by definition, and all the values correspond to $\omega_2=10^{-3}$ and $k'\sim M$.}
    \label{tab:table1}
\end{table}

An estimate of the tensor-to-scalar ratio ($r$) also follows immediately from the relation $r\sim16|\epsilon(N_{k_*})|$ \cite{Chung:2003iu}. We have $|\epsilon(k_*)|\sim10^{-11}$ leading to a generic estimate of $r\sim10^{-10}$, which is well consistent with the typical upper bound of $r\lesssim3\times10^{-2}$ obtained from joint analyses of the Planck 2018 and latest BICEP/Keck datasets \cite{BICEP:2021xfz,Tristram:2021tvh,Campeti:2022vom}. This could also be understood in terms of the Lyth bound, which renders $r$ the order of $(\hat{V}_{,c}/\hat{V})$ \cite{Lyth:1996im,RevModPhys.69.373,Efstathiou:2005tq,Easther:2006qu}.

Before concluding this section, we would like to emphasize the point that these values are merely representative ones for particular combinations of parameters, which have been chosen to demonstrate that the model is capable of successfully driving inflation while remaining consistent with currently available data. Any stronger comment on these estimates, however, demands an exhaustive scanning of the entire parameter space, which is not the focus of this study. The key takeaway from this section is the fact that in the curved doubly warped scenario, the modulus $k$ associated with the second extra dimension can play the role of an interesting degree of freedom which can parametrically tune the height (and the slope) of the inflationary potential across several orders of magnitude, with insignificant impact on the observable quantities of interest. 

\subsection{Inflation driven by $k$} \label{subsec:kinf}
\begin{figure}[!t]
    \centering
    \subfigure[]{\includegraphics[width=0.48\textwidth]{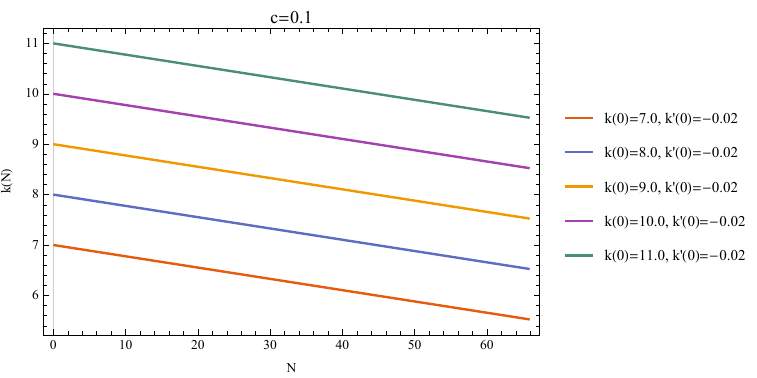}}
    \hfill
    \subfigure[]{\includegraphics[width=0.48\textwidth]{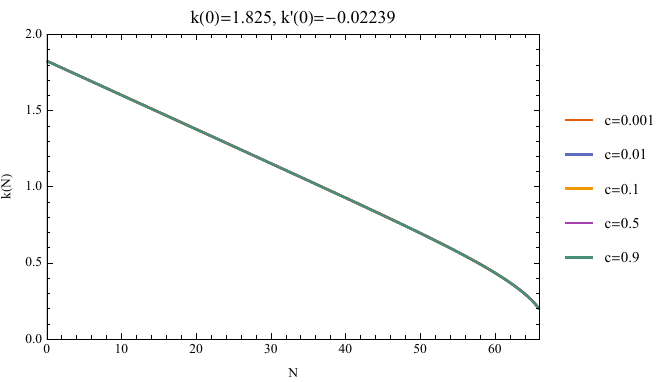}}
    \subfigure[]{\includegraphics[width=0.48\textwidth]{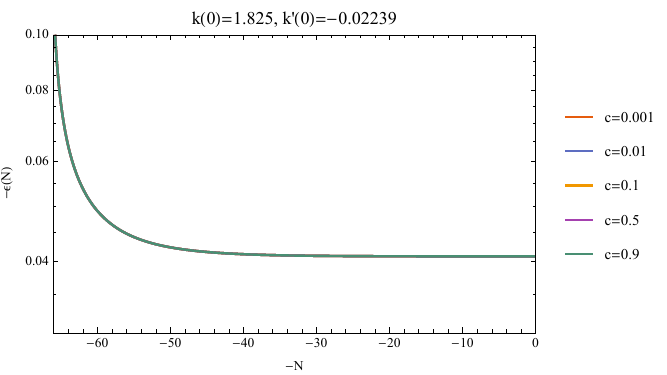}}
    \hfill
    \subfigure[]{\includegraphics[width=0.48\textwidth]{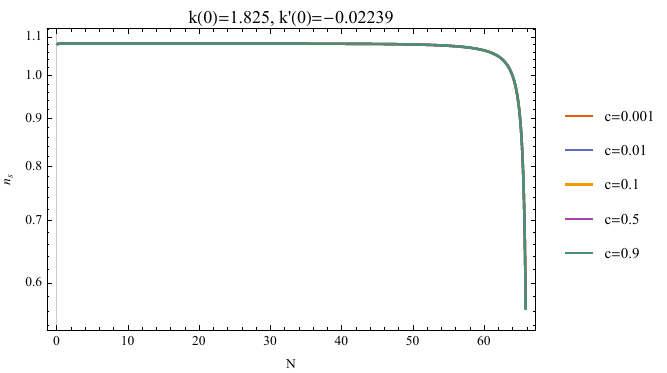}}
    \caption{Representative plots of $k(N)$, $\epsilon(N)$, and $n_s(N)$ for a few benchmark values of $c$, and a few particular initial condition imposed on $k$. For different initial conditions, the $k(N)$ curves run parallel to each other, leading eventually to different values of $N_{\rm tot}$ (upper left). For a given initial condition, the trajectories are independent of the value of $c$ and coincide with each other (upper right). The initial conditions chosen for the upper right panel result in exit from inflation around $N_{\rm tot}\sim66-70$. But $n_s(N)$ turns out to be greater than unity throughout most of inflation, except very close to the exit where it drops sharply. Note that in the bottom-left plot, inflation proceeds from right to left, and the steep upward curve towards the left margin actually represents the divergence of $|\epsilon(N)|$ close to that point signifying the the end of inflation. While extending the vertical range of the plot might better illustrate this divergence, it visually flattens the rest of the curve and makes it difficult to show the precise evolution of $\epsilon(N)$ throughout the course of inflation. All the plots correspond to $\omega_2=10^{-3}$ and $k'\sim M$.}
    \label{fig:k3_exact}
\end{figure}

In this section, we focus on the opposite scenario, \textit{i.e.} we treat $c$ as a parameter and consider the variation of \eqref{eq:seff} w.r.t. $k$ only. Similar to the previous section, using the Friedmann equations in the equation of motion for $k$, the latter can be recast in dimensionless form as
\begin{equation} \label{eq:keqexact}
    2\hat{\tilde{A}}_{kk}\left(\underaccent{\boldsymbol{\cdot}\boldsymbol{\cdot}}{k}+3\underaccent{\boldsymbol{\cdot}}{k}+\dfrac{1}{4}\hat{\tilde{A}}_{kk}\underaccent{\boldsymbol{\cdot}}{k}^3\right)+\hat{\tilde{A}}_{kk,k}\underaccent{\boldsymbol{\cdot}}{k}^2=12\left(\dfrac{\hat{V}_{,k}}{\hat{V}}\right)\left(1+\dfrac{1}{12}\hat{\tilde{A}}_{kk}\underaccent{\boldsymbol{\cdot}}{k}^2\right)\:.
\end{equation}
In this case, the ratio $(\hat{V}_{,k}/\hat{V})$ turns out to be independent of the value of $c$, as shown in Fig. \ref{fig:app1}. The slow roll parameter is given by
\begin{equation} \label{eq:epsk}
    \epsilon(N)\equiv-\dfrac{\dot{H}}{H^2}=-\dfrac{1}{4}\hat{\tilde{A}}_{kk}\underaccent{\boldsymbol{\cdot}}{k}^2\:\:\xrightarrow[]{\text{slow roll}}\:\:-\dfrac{1}{\hat{\tilde{A}}_{kk}}\left(\dfrac{\hat{V}_{,k}}{\hat{V}}\right)^2.
\end{equation}
Unlike the previous case, there is no inflection point here on a constant $c$-slice of the potential, which decays monotonically with increasing $k$. Consequently, one expects more or less qualitatively uniform behaviour with respect to the initial conditions $k(0)$ and $\underaccent{\boldsymbol{\cdot}}{k}(0)$. However, this scenario is interesting because the height of $\hat{V}(c,k)$ may vary appreciably over the course of inflation, \textit{i.e.} the potential can be steep. This does not necessarily spoil slow roll in non-canonical frameworks, as long as the non-canonicity allows $\epsilon(N)$ to remain smaller than unity \cite{Anber:2009ua,Adshead:2012kp,Rezazadeh:2015dia,Adshead:2016iix}. In fact, using \eqref{eq:epsk}, one may immediately arrive at a rough estimate of $|\epsilon(N)|\sim\mathcal{O}(10^{-2})$ based on Fig. \ref{fig:adsset} and Fig. \ref{fig:app1}, which show respectively that $\hat{\tilde{A}}_{kk}\sim300$ and $|(\hat{V}_{,k}/\hat{V})|\sim10$ for $k\gtrsim1$. Evidently, $\epsilon(N)$ diverges around $k\approx0.14$, where the denominator $\hat{\tilde{A}}_{kk}$ vanishes. Now, for initial $k(0)>0.14$, the kinetic term is non-canonical, which makes the field roll up the potential slope towards progressively smaller values of $k$. This makes it eventually reach the critical value of $k\approx0.14$, where inflation comes to an end. Hence, similar to the $c$-driven case, this scenario is also capable of driving slow roll inflation, besides being equipped with an in-built exit mechanism. These properties are demonstrated in the upper panel of Fig. \ref{fig:k3_exact} for a particular set of values. 

However, checking the viability of this scenario requires evaluating the scalar spectral index, which can be obtained as before as
\begin{equation} \label{eq:nsk}
    n_s(N)\:\approx\:1-3\left(1+\dfrac{p_k}{\rho_k}\right)-\dfrac{1}{H}\dfrac{d}{dt}\left[\textrm{ln}\left(1+\dfrac{p_k}{\rho_k}\right)\right]=1+\dfrac{1}{2}\hat{\tilde{A}}_{kk}\underaccent{\boldsymbol{\cdot}}{k}^2-\left(\dfrac{\hat{\tilde{A}}_{kk,k}}{\hat{\tilde{A}}_{kk}}\right)\underaccent{\boldsymbol{\cdot}}{k}-2\left(\dfrac{\underaccent{\boldsymbol{\cdot}\boldsymbol{\cdot}}{k}}{\underaccent{\boldsymbol{\cdot}}{k}}\right)\:,
\end{equation}
where $p_k$ and $\rho_k$ are defined in an analogous manner as earlier, and \eqref{eq:nsk} needs to be evaluated at horizon crossing of the relevant mode. We find that the scalar spectral index apparently remains fixed at a value of $n_s(N)\approx1.082$ throughout most of the inflationary phase, quasi-independently of both the initial condition and the value of $c$. Close to the exit point, it plunges rapidly and seemingly passes through the observed CMB bounds. But the corresponding number of $e$-folds from the end of inflation where this happens is very small, and the associated scale is consequently expected to be much shorter than the CMB pivot scale $k_*=0.05$ Mpc\textsuperscript{-1}. On the other hand, at the horizon-crossing $e$-fold of the CMB scale, $n_s(k_*)$ remains close to $1.08$. While one may hope to increase $N_{\rm tot}$ by raising $k(0)$ (\textit{viz.} Fig. \ref{fig:k3_exact}), with the aim of consistently matching the number of $e$-folds from the end of inflation where $n_s\approx0.96$ with the scale $k_*=0.05$ Mpc\textsuperscript{-1}, this idea is rather unappealing. This seems to require a much larger $k(0)$ than the values considered in Fig. \ref{fig:k3_exact}, which in turn would require an extremely small value of $c$ in order to produce a moderate $T_1/T_2$ hierarchy during the initial stages of inflation. Then, as $k(N)$ decreases quasi-linearly and reaches $k(N_*)\lesssim\mathcal{O}(1)$ at the tentative moment of horizon crossing, the same value of $c$ as before would now produce a very tiny $(T_1/T_2)_*$, which would imply that $T_2$ is extremely large compared to $T_1$. A tension thus emerges between early and late time values of $T_1/T_2$ during inflation, as it appears impossible to keep this ratio within moderate bounds throughout the entire duration of inflation due to the underlying geometry, if one attempts to make this scenario compatible with observations.

Since $|\epsilon(N)|\sim10^{-2}$ is not very small in this scenario, it makes sense to ask whether going to higher orders in $\epsilon(N)$ may significantly alter these conclusions and help drag the value of $n_s(k_*)$ towards the viable range. To investigate this, we calculate the correction $\Delta n_s$ incurred by a higher order term as follows. The scalar spectral index is defined strictly in terms of the scalar power spectrum ($P_k^\zeta$) as
\begin{equation}
    n_s-1\equiv\dfrac{d\textrm{ln}P_k^\zeta}{d\textrm{ln}k}=\dfrac{d}{d\textrm{ln}k}\textrm{ln}\left[\dfrac{16}{9}\left(\dfrac{\rho}{M}\right)\left(1+\dfrac{p}{\rho}\right)^{-1}\right]\:.
\end{equation}
The horizon-crossing condition imposes $d\textrm{ln}k=d\textrm{ln}a+d\textrm{ln}H$, where a nearly constant value of the Hubble parameter during inflation enables one to drop the second term and approximate the derivative operator as $d/d\textrm{ln}k\approx H^{-1}d/dt$ \cite{Enqvist:2012qc}. Combining the result with the continuity equation $\dot{\rho}+3H(\rho+p)=0$, one then ends up with the expressions of $(n_s-1)$ in \eqref{eq:nsc} and \eqref{eq:nsk}. However, motivated by the steepness of $\hat{V}$ in the present case, if one is to retain the $d\textrm{ln}H$ term, then up to first order, the derivative operator at horizon-crossing becomes $d/d\textrm{ln}k=H^{-1}(1+H^{-2}\dot{H})^{-1}d/dt\approx H^{-1}(1-H^{-2}\dot{H})d/dt$. This extra term results in the correction $\Delta n_s$ on top of the $n_s$ value calculated earlier, and is given by
\begin{equation}
    \Delta n_s\equiv\epsilon(N)(n_s^{(0)}-1)\:,
\end{equation}
where $n_s^{(0)}$ is the earlier result, and we have used the definition $\epsilon(N)=-\dot{H}/H^2=-\underaccent{\boldsymbol{\cdot}}{H}/H$. Recall that $n_s^{(0)}>1$ and $\epsilon(N)<0$, which encouragingly render $\Delta n_s<0$ and indicate the possibility of lowering the value of $n_s^{(0)}$. But the magnitude of this correction, calculated using $n_s^{(0)}\approx1.082$ and $\epsilon(N)\approx-0.04$, turns out to be $\Delta n_s\approx-0.003$, which is far too small compared to what is required to pull the value of $n_s$ down into the observational range. Hence, our remark about the unfeasible nature of the $k$-driven regime seems robust against higher-order slow-roll corrections. As future works, it may be interesting to study whether any additional spectator(s) and/or higher-dimensional flux field(s) could furnish a larger correction and help make this a viable scenario.

\section{Generic trends in higher dimensions} \label{sec:gentrends}
\begin{figure}[!t]
    \centering
    \subfigure[]{\includegraphics[width=0.3\textwidth]{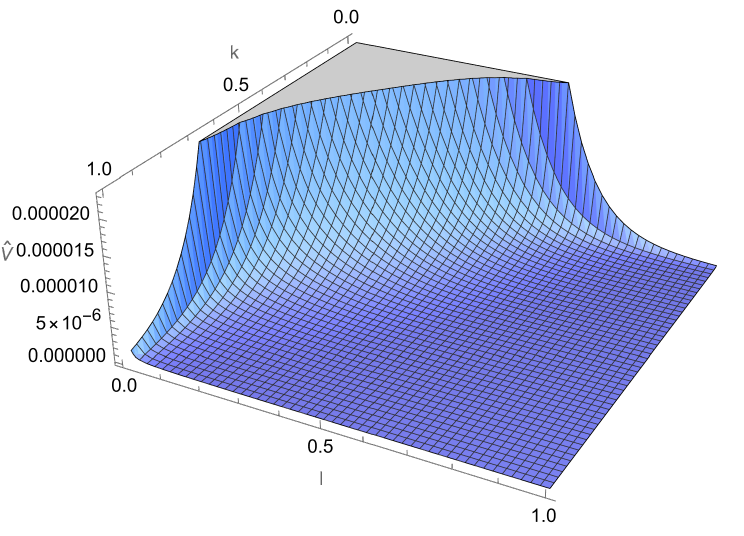}}
    \subfigure[]{\includegraphics[width=0.3\textwidth]{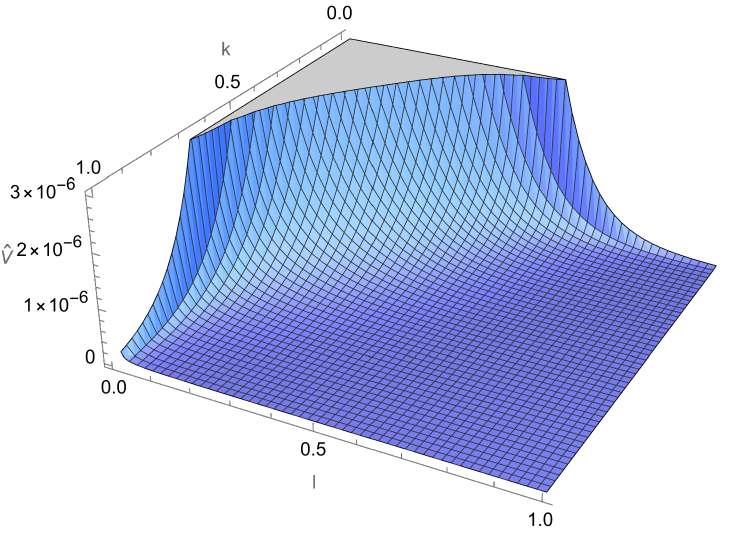}}
    \subfigure[]{\includegraphics[width=0.3\textwidth]{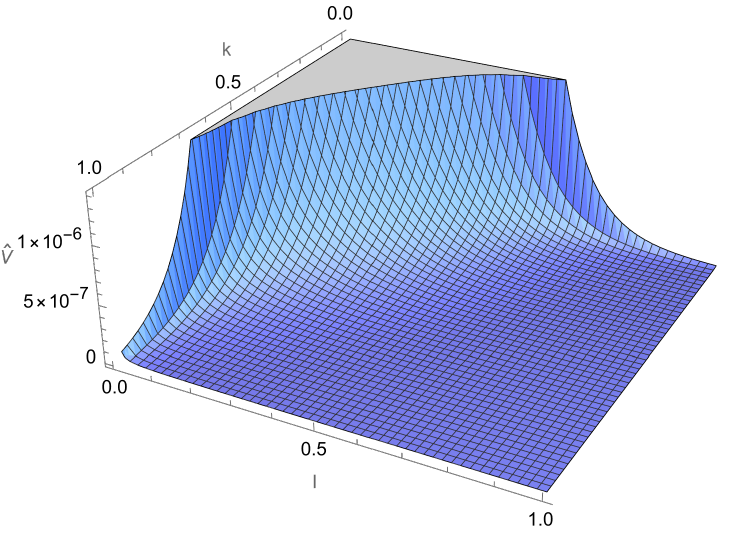}}
    \subfigure[]{\includegraphics[width=0.3\textwidth]{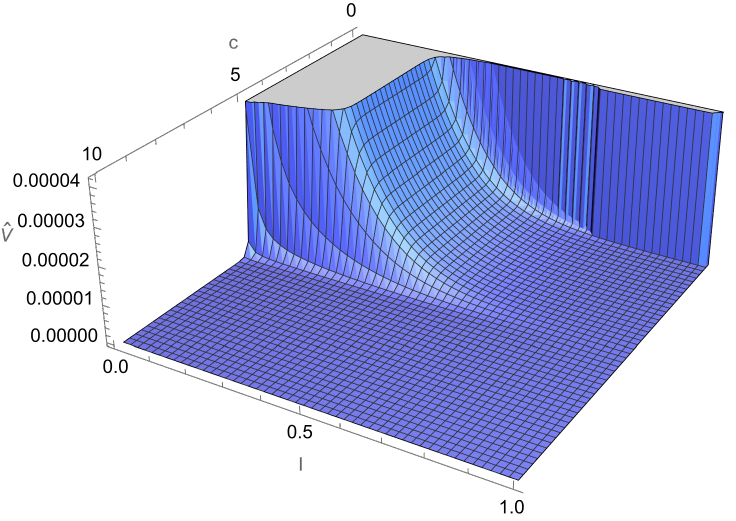}}
    \subfigure[]{\includegraphics[width=0.3\textwidth]{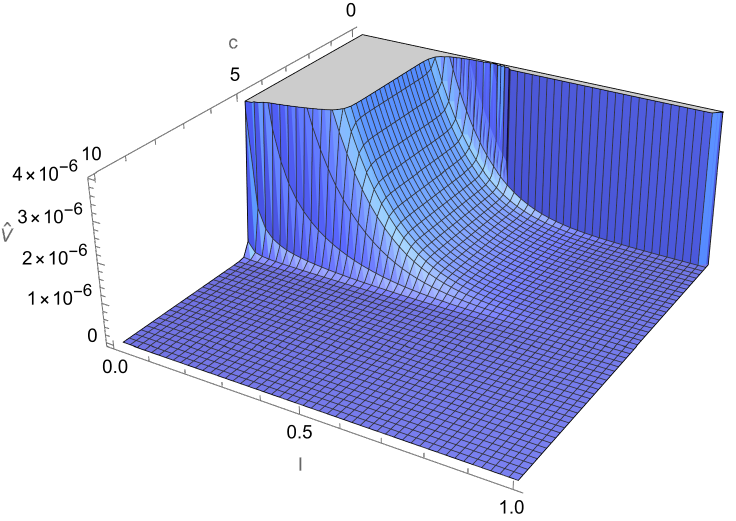}}
    \subfigure[]{\includegraphics[width=0.3\textwidth]{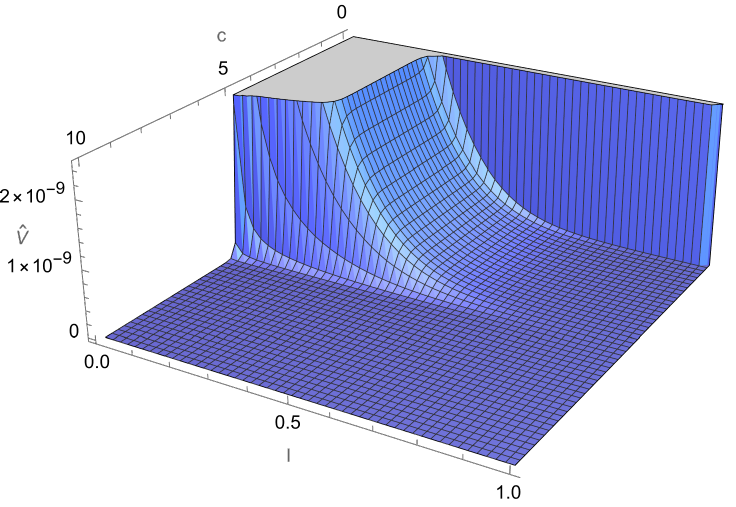}}
    \subfigure[]{\includegraphics[width=0.3\textwidth]{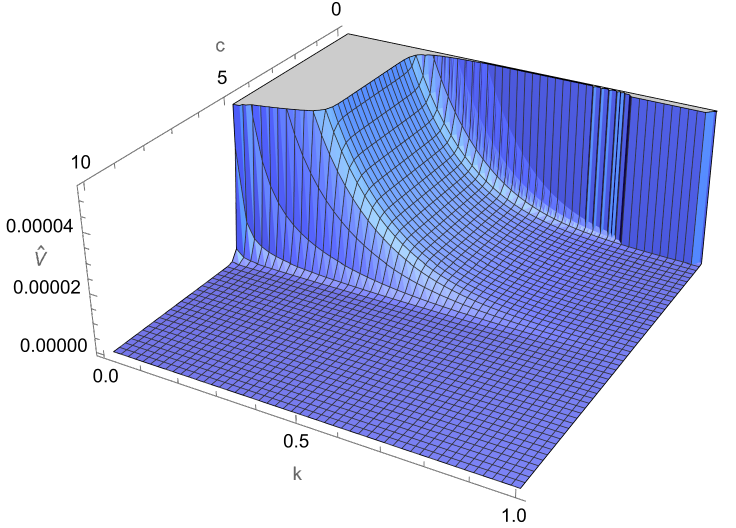}}
    \subfigure[]{\includegraphics[width=0.3\textwidth]{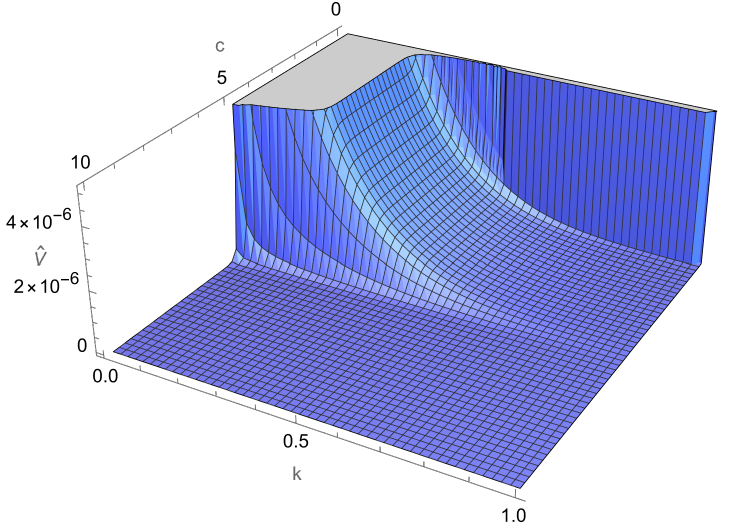}}
    \subfigure[]{\includegraphics[width=0.3\textwidth]{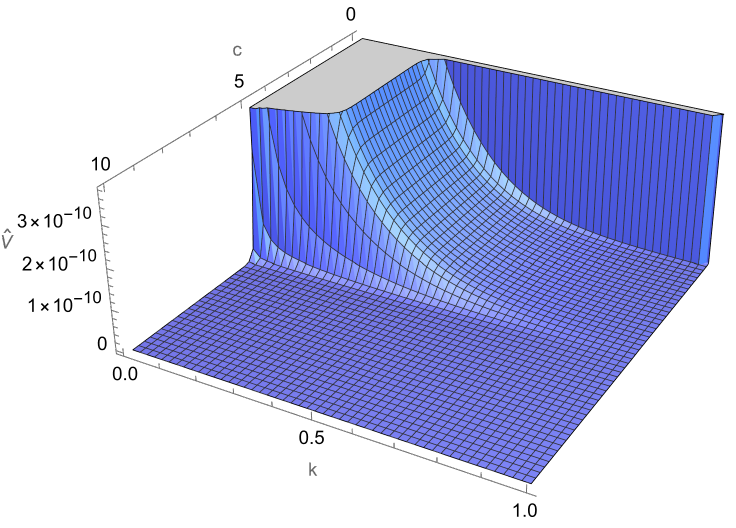}}
    \caption{Visual projections of the 4D dS ($\omega_2=10^{-3}$) effective potential $\hat{V}(c,k,\ell)$ along each modular direction as a function of the other two for a curved 7D nested braneworld. \textit{Top panel:} Constant $c$-slices at the values $c=0.01,\:0.1,\:1.0$. \textit{Middle panel:} Constant $k$-slices at the values $k=0.01,\:0.1,\:1.0$. \textit{Bottom panel:} Constant $\ell$-slices at the values $\ell=0.01,\:0.1,\:1.0$. Although the projections give the impression that $\hat{V}(c,k,\ell)$ is symmetric in $k$ and $\ell$, it is not exactly so due to the nested geometric structure which fundamentally places $k$ and $\ell$ on unequal footings. Note that all the quantities involved in the figure above are dimensionless by definition.}
    \label{fig:pots7d}
\end{figure}

In this section, we briefly highlight a few characteristics that appear to be generic dynamical features of non-flat braneworlds with arbitrary levels of nested warpings. For the purpose of demonstration, we first focus on the 4D de Sitter regime of the curved triply warped scenario. This scenario involves three moduli which we denote by $c$, $k$, and $\ell$. Using the associated warp factors and brane tensions derived in \cite{Bhaumik:2023tmg}, the Einstein frame potential $\hat{V}(c,k,\ell)$ can be derived following the procedure of Sec. \ref{sec:modpot}. We do not provide the full analytical form here due to its cumbersome appearance. Instead, we show its nature graphically in Fig. \ref{fig:pots7d} by plotting its projections along the three moduli separately. It is easy to see that the projections on constant $k$ and $\ell$-slices look very similar to each other, which may be traced back to the identical functional form of the warp factors along the corresponding orbifolds. This is also apparent in the constant $c$ projections. However, the presence of nested warping puts $k$ and $\ell$ on fundamentally different footings, which induces small differences in the behaviour of the resulting $\hat{V}(c,k,\ell)$ along these two moduli, thus preventing them from being exactly identical to each other from a numerical point of view. 

In a similar manner, the non-canonical kinetic couplings of the 4D effective action may also be derived in the Einstein frame. In this case, a total of six self and cross-kinetic terms arises among the three moduli fields, leading to six coupling functions which we denote by $\hat{\tilde{A}}_{cc}$, $\hat{\tilde{A}}_{kk}$, $\hat{\tilde{A}}_{\ell\ell}$, $\hat{\tilde{A}}_{ck}$, $\hat{\tilde{A}}_{k\ell}$, and $\hat{\tilde{A}}_{c\ell}$. In the spirit of Sec. \ref{sec:inflation}, we are more interested here in the first three. It turns out that $\hat{\tilde{A}}_{cc}(c)$ and $\hat{\tilde{A}}_{kk}(k)$ are completely identical to their 6D counterparts and receive no direct modification from the additional level of warping. On the other hand, the shape of $\hat{\tilde{A}}_{\ell\ell}(\ell)$ turns out to be very similar to that of $\hat{\tilde{A}}_{kk}(k)$, with what appears to be effectively an overall rescaling alongside a minor lateral shift (\textit{viz.} upper left panel of Fig. \ref{fig:dsset}). We find that $\hat{\tilde{A}}_{\ell\ell}(\ell)\to-\infty$ as $l\to0$, and $\hat{\tilde{A}}_{\ell\ell}(\ell)\sim630$ for $\ell\gtrsim1$, with a zero-crossing around $\ell\approx0.12$. These features are independent of the value of the induced cosmological constant, as, like $\hat{\tilde{A}}_{kk}(k)$, $\hat{\tilde{A}}_{\ell\ell}(\ell)$ is also independent of $\omega_2$. Qualitatively similar features are found to arise in the AdS scenario as well.

Equipped with these results, we are now in a position to make some generic comments. Firstly, for an $n$-fold nested warped geometry with $\Omega\neq0$, the induced potential $\hat{V}$ in the Einstein frame (that vanishes identically in the $\Omega\to0$ limit) exhibits a metastable minimum in the direction of the first extra-dimensional modulus, and appears to decay monotonically along all the higher modular directions. The functional dependences of $\hat{V}$ on these higher moduli are expected to be very similar to each other with minute differences, owing to the underlying nested geometry and the identical functional forms of the corresponding warp factors. Besides the potential, there arise a total of $n+{n\choose2}$ non-canonical moduli kinetic terms in the 4D effective action. In the Einstein frame, the non-canonical coefficient of each self-kinetic term is a function only of the modulus of that term, while that of each cross-kinetic term depends on the two moduli corresponding to that term. Moreover, only the first modulus remains explicitly associated with $\Omega$, which does not show up in any of the kinetic coefficients that does not involve at least one derivative of the first modulus. On the other hand, the coefficients of the higher-moduli kinetic terms (both self and cross) exhibit features very similar to each other as functions of their corresponding moduli, which is once again due to the similarity of the higher warp factors. From a cosmological point of view, predominant slow roll on the metastable plateau along the first modulus provides a viable inflationary mechanism with a natural exit scheme, both concordant with current observations. Interestingly, the inflationary parameters remain largely independent of the values of the other moduli, whose key role consists of regulating the height of the potential plateau. On the other hand, while slow roll scenarios along the higher moduli can produce an inflating background alongside a proper exit, they are disfavoured as good agreement with data seemingly demands a huge variation of the extra-dimensional size hierarchy over the course of inflation.

\section{Discussions and conclusion} \label{sec:disc}

In the present work, we have analyzed the 4D effective dynamics of extra-dimensional moduli fields in the backdrop of nested warped geometry with curved branes, with emphasis on their cosmological prospects for an inflationary epoch that is consistent with observations. The whole analysis has been carried out explicitly in terms of the non-canonical moduli fields themselves, instead of what is conventionally done in terms of the canonically normalized radion field(s). The former approach allows greater physical insight into our results directly in terms of the geometric properties of the underlying braneworld, which may be somewhat obscured in the latter one. In this section, we summarize our key findings in order, and briefly comment on a few interesting future directions to explore.

Following a brief recap of the essential geometric features of the curved doubly warped model in Sec. \ref{sec:review}, we have started in Sec. \ref{sec:modpot} by explicitly deriving the 4D induced potential and the modular kinetic terms in the Jordan frame, which involves a non-minimal coupling of the 4D Ricci scalar to the moduli fields themselves. This has been achieved by splitting the total 6D Ricci scalar appropriately and identifying the specific type of effective lower-dimensional term arising from each part. The derivatives with respect to the extra-dimensional angular coordinates, alongside the bulk cosmological constant and the 4-brane tension terms in the full 6D action, lead to the effective potential $V(c,k)$, which shows distinct behaviour in the dS and AdS regimes (Sec. \ref{subsec:modpotorig}). The derivatives with respect to the usual non-compact 4D coordinates, on the other hand, lead to the effective kinetic terms after suitable integration by parts, whereas the non-minimal Jordan frame coupling emerges from the fact that the 6D Ricci scalar contains the warped 4D Ricci scalar within itself (Sec. \ref{subsec:curvkin}). The entire Jordan frame action has then been recast in the Einstein frame in Sec. \ref{subsec:jordeinst}, where the non-minimal gravitational coupling is removed at the cost of non-trivial modifications (involving the same coupling function) to the form of the effective potential, as well as to those of the kinetic terms. For $\Omega>0$ in particular, which is the case of physical interest, the Einstein frame potential $\hat{V}(c,k)$ contains a metastable plateau of finite width in the direction of the first modulus $c$, while decaying with no minima along the second modulus $k$. Thus, it appears that positively curved brane geometry alone cannot stabilize all of the moduli in a metastable fashion in multiply warped settings, and it becomes important to invoke additional stabilizing mechanisms for the higher moduli. On the other hand, the non-canonical kinetic couplings in the Einstein frame are found to depend on the moduli following a nice pattern, with the coupling function of each self-kinetic term depending only on the corresponding modulus, and that of the single cross-kinetic term depending on both $c$ and $k$. This is but a consequence of the nested geometric structure, which allows such a splitting in the Einstein frame which is equipped with a minimal gravitational coupling. Moreover, the input of the brane curvature appears in only those kinetic terms which involve at least one derivative of $c(x)$. This, in turn, is a consequence of the fact that only the first warp factor $a(y)$ (which is associated with $c$) is explicitly dependent on $\Omega$, which does not show up in the expression of any subsequently higher warp factor. As we demonstrate by means of a 7D extension in Sec. \ref{sec:gentrends}, most of these dynamical features can be generalized readily to any arbitrary $n$-fold warped braneworld with $n\geq2$ in presence of non-zero brane curvature. Furthermore, in Sec. \ref{subsec:fieldcurv}, we have explicitly computed the field space curvature for the non-canonical kinetic scenario that emerges in the Einstein frame, in both the AdS and the dS regimes separately. This analysis supplements the preceding calculations in a more robust manner, and illustrates interesting trends in the variation of non-canonicity as a function of the dimensionless moduli fields $c$ and $k$.

In Sec. \ref{sec:inflation}, we have made use of the results derived in Sec. \ref{sec:modpot} to study the prospects for slow roll inflation, which exemplifies a physical scenario where the idea of such a curved braneworld may prove particularly useful. While a full-fledged two-field analysis involving both of the moduli is possible in principle, it is numerically expensive and not entirely relevant within the framework of the present study, which primarily aims to investigate, as part of its full scope, some possible cosmological phenomenon which might be addressed naturally from the perspective of a non-flat multiply warped braneworld. As such, it would be interesting to pursue the former route in future as part of a detailed study focused purely on braneworld-induced multi-field inflation. In this work, we have restricted ourselves to a case-by-case analysis instead, where we have explored the consequences of slow roll along the potential predominantly in one particular modular direction at a time, with the other modulus held approximately fixed and thus playing a parametric role. While we maintain an agnostic stance towards any possible process(es) which may absolutely stabilize the parametric modulus, we find that it is nonetheless possible to achieve a nearly constant value of any one modulus field while allowing the other one to roll, in specific regions of the field space in our scenario. Besides, in fundamentally string-inspired frameworks which may give rise to effective warped geometries in lower dimensions, it is not altogether impossible to imagine more sophisticated moduli stabilizing processes resulting from $p$-form fluxes on the higher-dimensional Calabi-Yau manifolds \cite{Verlinde:1999fy,Chan:2000ms,Kachru:2002he,Brummer:2005sh}. We have not pursued this line of inquiry any further and left it to a future study, focusing instead on the physically observable consequences of such unidirectional slow-roll scenarios.

To that end, we have investigated inflationary scenarios driven individually by $c(x)$ and $k(x)$ in Sec. \ref{subsec:cinf} and Sec. \ref{subsec:kinf} respectively. Firstly, it is interesting to note that both of the cases come equipped with in-built exit mechanisms at some finite $e$-fold value, which is linked to the nature of the non-canonical kinetic coefficients $\hat{\tilde{A}}_{cc}(c)$ and $\hat{\tilde{A}}_{kk}(k)$. Each of these coefficients exhibits a zero-crossing in its corresponding field space, where the kinetic term switches its sign. For suitable initial values greater than this critical value, the kinetic term is phantom, and the field slowly climbs up the potential towards this tipping point. The slow roll parameter, being inversely proportional to the non-canonical kinetic coefficient, eventually diverges at this point, marking an exit from inflation. We would like to emphasize that our scenario is distinct from other inflationary proposals involving negative kinetic terms, \textit{\textit{e.g.}} phantom inflation \cite{Piao:2004tq,Liu:2012iba,Richarte:2016qqm} or ghost inflation \cite{Arkani-Hamed:2003juy,Jazayeri:2016jav,Ivanov:2014yla}, due to the simultaneous presence of the non-trivial kinetic coefficients and the effective potential, neither of which needs to be invoked \textit{ad hoc} in our case. 

Depending on the initial conditions, both the $c$-driven and the $k$-driven cases are found capable of generating a sufficient number of $e$-foldings as required observationally. However, in light of the latest bounds on the scalar spectral index, we find the $c$-driven case to be the more favoured scenario between the two. For the benchmark value of $\omega_2=10^{-3}$, we find that good agreement with data requires a total of $\mathcal{O}(10^3)$ $e$-folds (of which the last $\sim60-70$ correspond to physical inflation), for an initial value $c(0)$ tuned close to the inflection point on a constant $k$-slice of $\hat{V}(c,k)$. For smaller values of $\omega_2$, a finer tuning of the initial condition may be necessary. The value of $k$ does not seem to impact the inflationary observables significantly, but is capable of modulating the height of the potential across several orders of magnitude while maintaining a moderate hierarchy between the radii of the extra dimensions. This flexibility might be interesting from the perspective of theoretical arguments for a reduced energy scale of inflation, \textit{\textit{e.g.}} based on the trans-Planckian censorship conjecture (TCC) and other swampland criteria (SC) \cite{Brennan:2017rbf,Palti:2019pca,Bedroya:2019tba,Brandenberger:2021pzy}. 

In the $k$-driven case, on the other hand, $c$ plays a similar parametric role to tune the height of the potential. But in this case, our results indicate a slight blue tilt of the scalar power on CMB scales, for moderate choices of the initial conditions on $k(x)$ and values of $c$. This is incompatible with the latest CMB observations, which report a weak red tilt instead. While our analysis hints that it might be possible to ameliorate this issue by increasing the total number of $e$-folds, this possibility corresponds to $k(0)\gtrsim\mathcal{O}(10)$, which requires a very small value of $c\sim e^{-k(0)\pi}$ to control the hierarchy $T_1/T_2$ between the extra-dimensional radii during the initial moments of inflation. However, as the exit occurs near $k\sim0.1$, one ends up with $T_2\gg T_1$ close to the end of inflation, which spells trouble for the size hierarchy. On the other hand, if one chooses to have a moderate $T_1/T_2$ ratio close to the end instead, then initially one must have $T_1\gg T_2$. Compatibility of the $k$-driven inflationary regime with observations thus apparently requires a drastic fall in the $T_1/T_2$ ratio over the course of inflation, with at least one end ending up with an alarmingly large extra dimension. Based on similar trends of the potential and the kinetic coefficients in higher-dimensional extensions of the 6D model, we infer in Sec. \ref{sec:gentrends} that these may be generic features of slow roll along any of the higher moduli beyond the first one. While this compels us to discard this regime within the purview of the present work, it might be interesting to study in future whether simultaneous variation of $c$, or even non-trivial stabilization mechanisms which may possibly fix the value of $c$, might help make this a viable scenario. 

In the end, we would like to note that the present analysis is by no means exhaustive, and that there may be numerous interesting extensions. Reiterating our comment from earlier, the first avenue to explore could be whether it is possible to obtain a concrete stabilization mechanism which may justify the unidirectional slow roll scheme and the stability of the other moduli more rigorously, which we have explored as particular cases of interest in this analysis. This could be studied, for instance, in the context of Type IIB string-inspired frameworks where fluxes play an important stabilizing role. The second avenue would be to proceed with a full numerical computation of both the background and perturbations by keeping all the moduli free, thus approaching the scenario from the perspective of multi-field inflation which we have deliberately avoided in this work. Apart from these two, other possibilities include studying late time cosmological effects in the dS branch, or potential physical applications of the AdS branch, which have been relatively less explored even in the 5D scenario. We plan to report on some of these aspects in future works. 

\acknowledgments

The research work of AB is supported by CSIR Senior Research Fellowship (File no. \hash{09/0093(13641)/2022-EMR-I}).

\appendix

\section{List of explicit functional forms} \label{sec:appendixA}

In this appendix, we provide a list of the full expressions of the conformal coupling factor $h(c,k)$, the Einstein frame moduli potential $\hat{V}(c,k)$, and the non-canonical kinetic coefficients in the Einstein frame, for both the AdS and the dS branches in the curved doubly warped geometry. While our actual calculations have proceeded from the Jordan frame to the Einstein frame, the latter has been central to our analysis. This is why we present only the Einstein frame quantities here. For computational purposes, it suffices to recall that the Jordan and Einstein frame potentials are related via $\hat{V}=V/h^2$, and the analogous relations for the kinetic coefficients are given by \eqref{eq:kinrels}. These relations fully specify the Jordan frame quantities in terms of the Einstein frame, whenever required. 

\subsection{Anti-de Sitter}

\begin{itemize}
\item \textit{Conformal coupling factor}:
{
 \allowdisplaybreaks
  \begin{align} \label{appeq:adsh}
        h(c,k)=& -\frac{1}{12} \omega_1^2\text{sech}^2(\pi  k) \left[9 \sinh (\pi  k)+\sinh (3 \pi  k)\right] \nonumber \\
        & \times \Bigg[\sinh \left[2\pi  c+2\ln \left(\frac{\omega_1}{c_1}\right)\right]+2 \pi  c-\sinh \left[2 \ln \left(\frac{\omega_1}{c_1}\right)\right]\Bigg]
  \end{align}
}
\item \textit{Moduli potential (Einstein frame)}:
{
 \allowdisplaybreaks
  \begin{align} \label{appeq:adsmodpot}
            \hat{V}(c,k)=\:\:& 12 \sinh (\pi  k) \left[c_1^4 \omega_1^4 (9 \sinh (\pi  k)+\sinh (3 \pi  k))^2 \left\{\sinh \left(2 \left(\pi  c+\log \left(\frac{\omega_1}{c_1}\right)\right)\right) \right. \right. \nonumber \\
            & \left. \left. +2 \pi  c-\sinh \left(2 \log \left(\frac{\omega_1}{c_1}\right)\right)\right\}^2\right]^{-1} \times \left[-6 c_1^4 \omega_1^4 \sinh \left(2 \left(\pi  (c+k)+\log \left(\frac{\omega_1}{c_1}\right)\right)\right) \right. \nonumber \\
            & -6 c_1^4 \omega_1^4 \sinh \left(2 \left(\pi  c+\log \left(\frac{\omega_1}{c_1}\right)-\pi  k\right)\right)-120 \pi  c c_1^4 \omega_1^4 \nonumber \\
            & -60 c_1^4 \omega_1^4 \sinh \left(2 \left(\pi  c+\log \left(\frac{\omega_1}{c_1}\right)\right)\right)+\cosh (2 \pi  k) \left\{6 c_1^4 \omega_1^4 \left(3 \sqrt{1-\omega_1^2}-4 \pi  c\right) \right. \nonumber \\
            & +c_1^8 \left(3 \sqrt{1-\omega_1^2}-4\right)+4 c_1^6 \omega_1^2 \left(3 \sqrt{1-\omega_1^2}-2\right)+4 c_1^2 \omega_1^6 \left(3 \sqrt{1-\omega_1^2}+2\right) \nonumber \\
            & \left. +\omega_1^8 \left(3 \sqrt{1-\omega_1^2}+4\right)\right\}+15 c_1^8 \sqrt{1-\omega_1^2}-15 c_1^8-60 c_1^6 \omega_1^2 \nonumber \\
            & +60 c_1^6 \omega_1^2 \sqrt{1-\omega_1^2}+4 c_1^4 \omega_1^4 \sinh \left(2 \left(\log \left(\frac{\omega_1}{c_1}\right)+\pi  k\right)\right) \nonumber \\
            & -c_1^4 \omega_1^4 \sinh \left(4 \log \left(\frac{\omega_1}{c_1}\right)+2 \pi  k\right)-4 c_1^4 \omega_1^4 \sinh \left(2 \pi  k-2 \log \left(\frac{\omega_1}{c_1}\right)\right) \nonumber \\
            & +c_1^4 \omega_1^4 \sinh \left(2 \pi  k-4 \log \left(\frac{\omega_1}{c_1}\right)\right)+90 c_1^4 \omega_1^4 \sqrt{1-\omega_1^2} \nonumber \\
            & \left. +60 c_1^2 \omega_1^6+60 c_1^2 \omega_1^6 \sqrt{1-\omega_1^2}+15 \omega_1^8+15 \omega_1^8 \sqrt{1-\omega_1^2}\right]
    \end{align}
}
\item \textit{Non-canonical kinetic coefficients (Einstein frame)}:
{
 \allowdisplaybreaks
  \begin{align} \label{appeq:adsacc}
            \hat{\tilde{A}}_{cc}(c)=\:& 4 \pi ^2 \left[\sinh \left(2\pi  c+2\log \left(\frac{\omega_1}{c_1}\right)\right)+2 \pi  c-\sinh \left(2 \log \left(\frac{\omega_1}{c_1}\right)\right)\right]^{-2} \nonumber \\
            & \times \Bigg[\left\{\sinh \left(2\pi  c+2\log \left(\frac{\omega_1}{c_1}\right)\right)+2 \pi  c-\sinh \left(2 \log \left(\frac{\omega_1}{c_1}\right)\right)\right\} \nonumber \\
            & \quad\quad\left\{3 \sinh \left(2\pi  c+2\log \left(\frac{\omega_1}{c_1}\right)\right)-2 \pi  c\right\}-12 \cosh ^4\left(\pi  c+\log \left(\frac{\omega_1}{c_1}\right)\right)\Bigg]
  \end{align}
}
{
 \allowdisplaybreaks
  \begin{align} \label{appeq:adsakk}
            \hat{\tilde{A}}_{kk}(k)=\:\:& \dfrac{3}{8 k^2} \text{sech}^2(\pi  k) \left[9 \sinh (\pi  k)+\sinh (3 \pi  k)\right]^{-2} \times \Big[-1407 \pi ^2 k^2+120 \pi ^2 k^2 \cosh (6 \pi  k) \nonumber \\
            & +11 \pi ^2 k^2 \cosh (8 \pi  k)+8 \left(41 \pi ^2 k^2+8\right) \cosh (2 \pi  k)-4 \left(51 \pi ^2 k^2+160\right) \cosh (4 \pi  k) \nonumber \\
            & +1344 \pi  k \sinh (2 \pi  k)+768 \pi  k \sinh (4 \pi  k)+64 \pi  k \sinh (6 \pi  k)-64 \cosh (6 \pi  k)+640\Big]
  \end{align}
}
{
 \allowdisplaybreaks
  \begin{align} \label{appeq:adsack}
            \hat{\tilde{A}}_{ck}(c,k)=\:\:&\dfrac{24 \pi  }{c} \text{sech}(\pi  k) \left[9 \sinh (\pi  k)+\sinh (3 \pi  k)\right]^{-1} \Bigg[\sinh \left(2\pi  c+2\log \left(\frac{\omega_1}{c_1}\right)\right)+2 \pi  c \nonumber \\
            & -\sinh \left(2 \log \left(\frac{\omega_1}{c_1}\right)\right)\Bigg]^{-1} \Bigg[\pi  c \cosh (4 \pi  k) \cosh \left(2\pi  c+2\log \left(\frac{\omega_1}{c_1}\right)\right) \nonumber \\
            & +2 \cosh ^4(\pi  k) \sinh \left(2\pi  c+2\log \left(\frac{\omega_1}{c_1}\right)\right)+3 \pi  c \cosh \left(2\pi  c+2\log \left(\frac{\omega_1}{c_1}\right)\right) \nonumber \\
            & +\sinh \left(2 \log \left(\frac{\omega_1}{c_1}\right)\right)+9 \pi  c \sinh ^2(\pi  k) \cosh \left(2\pi  c+2\log \left(\frac{\omega_1}{c_1}\right)\right) \nonumber \\
            & -\sinh (\pi  c) \cosh (4 \pi  k) \cosh \left(\pi  c+2 \log \left(\frac{\omega_1}{c_1}\right)\right) +\pi  c \sinh (\pi  k) \sinh (3 \pi  k) \nonumber \\
            & +5 \cosh (2 \pi  k) \left\{\pi  c \cosh \left(2\pi  c+2\log \left(\frac{\omega_1}{c_1}\right)\right) -\sinh (\pi  c) \cosh \left(\pi  c+2 \log \left(\frac{\omega_1}{c_1}\right)\right)\right\} \nonumber \\
            & +\pi  c \sinh (\pi  k) \sinh (3 \pi  k) \cosh \left(2\pi  c+2\log \left(\frac{\omega_1}{c_1}\right)\right) +9 \pi  c \sinh ^2(\pi  k)+\pi  c \Bigg]
  \end{align}
}
\end{itemize}

\subsection{de Sitter}
\begin{itemize}
\item \textit{Conformal coupling factor}:
{
 \allowdisplaybreaks
  \begin{align} \label{appeq:dsh}
        h(c,k)=& -\frac{1}{12}  \omega_2^2\text{sech}^2(\pi  k) \left[9 \sinh (\pi  k)+\sinh (3 \pi  k)\right] \nonumber \\
        & \times \Bigg[-\sinh \left[2 \pi  c-2 \ln \left(\frac{c_2}{\omega_2}\right)\right]+2 \pi  c-\sinh \left[2 \ln \left(\frac{c_2}{\omega_2}\right)\right]\Bigg]
  \end{align}
}
\item \textit{Moduli potential (Einstein frame)}:
{
 \allowdisplaybreaks
  \begin{align} \label{appeq:dsmodpot}
            \hat{V}(c,k)=\:\:& 12 \sinh (\pi  k)\left[c_2^4 \omega_2^4 (9 \sinh (\pi  k)+\sinh (3 \pi  k))^2 \left\{\sinh \left(2 \pi  c-2 \log \left(\frac{c_2}{\omega_2}\right)\right) \right. \right. \nonumber \\
            & \left. \left. -2 \pi  c+\sinh \left(2 \log \left(\frac{c_2}{\omega_2}\right)\right)\right\}^2\right]^{-1} \times \left[6 c_2^4 \omega_2^4 \sinh \left(2 \pi  (c-k)-2 \log \left(\frac{c_2}{\omega_2}\right)\right) \right. \nonumber \\
            & +6 c_2^4 \omega_2^4 \sinh \left(2 \pi  (c+k)-2 \log \left(\frac{c_2}{\omega_2}\right)\right)-120 \pi  c c_2^4 \omega_2^4 \nonumber \\
            & +60 c_2^4 \omega_2^4 \sinh \left(2 \pi  c-2 \log \left(\frac{c_2}{\omega_2}\right)\right)+\cosh (2 \pi  k) \left\{6 c_2^4 \omega_2^4 \left(3 \sqrt{\omega_2^2+1}-4 \pi  c\right) \right. \nonumber \\
            & +c_2^8 \left(3 \sqrt{\omega_2^2+1}-4\right)+4 c_2^6 \omega_2^2 \left(2-3 \sqrt{\omega_2^2+1}\right)-4 c_2^2 \omega_2^6 \left(3 \sqrt{\omega_2^2+1}+2\right) \nonumber \\
            & \left. +\omega_2^8 \left(3 \sqrt{\omega_2^2+1}+4\right)\right\}+15 c_2^8 \sqrt{\omega_2^2+1}-15 c_2^8+60 c_2^6 \omega_2^2 \nonumber \\
            & -60 c_2^6 \omega_2^2 \sqrt{\omega_2^2+1}+4 c_2^4 \omega_2^4 \sinh \left(2 \left(\log \left(\frac{c_2}{\omega_2}\right)+\pi  k\right)\right) \nonumber \\
            & +c_2^4 \omega_2^4 \sinh \left(4 \log \left(\frac{c_2}{\omega_2}\right)+2 \pi  k\right)-4 c_2^4 \omega_2^4 \sinh \left(2 \pi  k-2 \log \left(\frac{c_2}{\omega_2}\right)\right) \nonumber \\
            & -c_2^4 \omega_2^4 \sinh \left(2 \pi  k-4 \log \left(\frac{c_2}{\omega_2}\right)\right)+90 c_2^4 \omega_2^4 \sqrt{\omega_2^2+1} \nonumber \\
            & \left. -60 c_2^2 \omega_2^6-60 c_2^2 \omega_2^6 \sqrt{\omega_2^2+1}+15 \omega_2^8+15 \omega_2^8 \sqrt{\omega_2^2+1}\right]
  \end{align}
}
\item \textit{Non-canonical kinetic coefficients (Einstein frame)}:
    {
 \allowdisplaybreaks
  \begin{align} \label{appeq:dsacc}
            \hat{\tilde{A}}_{cc}(c)=& -2 \pi ^2 \left[\sinh \left(2 \pi  c-2 \log \left(\frac{c_2}{\omega_2}\right)\right)-2 \pi  c+\sinh \left(2 \log \left(\frac{c_2}{\omega_2}\right)\right)\right]^{-2} \nonumber \\
            & \times \Bigg[8 \pi ^2 c^2-4 \pi  c \sinh \left(2 \log \left(\frac{c_2}{\omega_2}\right)\right)+8 \pi  c \sinh \left(2 \pi  c-2 \log \left(\frac{c_2}{\omega_2}\right)\right) \nonumber \\
            & -12 \cosh \left(2 \pi  c-2 \log \left(\frac{c_2}{\omega_2}\right)\right)+3 \cosh \left(2 \pi  c-4 \log \left(\frac{c_2}{\omega_2}\right)\right) \nonumber \\
            & -3 \cosh (2 \pi  c)+12\Bigg]
  \end{align}
}
{
 \allowdisplaybreaks
  \begin{align} \label{appeq:dsakk}
            \hat{\tilde{A}}_{kk}(k)=& \dfrac{3}{8 k^2} \text{sech}^2(\pi  k) \left[9 \sinh (\pi  k)+\sinh (3 \pi  k)\right]^{-2} \times \Big[-1407 \pi ^2 k^2+120 \pi ^2 k^2 \cosh (6 \pi  k) \nonumber \\
            & +11 \pi ^2 k^2 \cosh (8 \pi  k)+8 \left(41 \pi ^2 k^2+8\right) \cosh (2 \pi  k)-4 \left(51 \pi ^2 k^2+160\right) \cosh (4 \pi  k) \nonumber \\
            & +1344 \pi  k \sinh (2 \pi  k)+768 \pi  k \sinh (4 \pi  k)+64 \pi  k \sinh (6 \pi  k)-64 \cosh (6 \pi  k)+640\Big]
  \end{align}
}
{
 \allowdisplaybreaks
  \begin{align} \label{appeq:dsack}
            \hat{\tilde{A}}_{ck}(c,k)=& -\dfrac{3 \pi}{c} \text{sech}(\pi  k)  \left[9 \sinh (\pi  k)+\sinh (3 \pi  k)\right]^{-1} \times \Bigg[\sinh \left(2 \pi  c-2 \log \left(\frac{c_2}{\omega_2}\right)\right)-2 \pi  c \nonumber \\
            & +\sinh \left(2 \log \left(\frac{c_2}{\omega_2}\right)\right)\Bigg]^{-1} \times \Bigg[6 \sinh \left(2 \pi  (c-k)-2 \log \left(\frac{c_2}{\omega_2}\right)\right) \nonumber \\
            & +6 \sinh \left(2 \pi  (c+k)-2 \log \left(\frac{c_2}{\omega_2}\right)\right)+\sinh \left(2 \pi  (c+2 k)-2 \log \left(\frac{c_2}{\omega_2}\right)\right) \nonumber \\
            & +\sinh \left(2 \pi  (c-2 k)-2 \log \left(\frac{c_2}{\omega_2}\right)\right)-24 \pi  c \cosh \left(2 \pi  (c-k)-2 \log \left(\frac{c_2}{\omega_2}\right)\right) \nonumber \\
            & -24 \pi  c \cosh \left(2 \pi  (c+k)-2 \log \left(\frac{c_2}{\omega_2}\right)\right)-3 \pi  c \cosh \left(2 \pi  (c+2 k)-2 \log \left(\frac{c_2}{\omega_2}\right)\right) \nonumber \\
            & -3 \pi  c \cosh \left(2 \pi  (c-2 k)-2 \log \left(\frac{c_2}{\omega_2}\right)\right)-6 \sinh \left(2 \pi  c-2 \log \left(\frac{c_2}{\omega_2}\right)\right) \nonumber \\
            & +54 \pi  c \cosh \left(2 \pi  c-2 \log \left(\frac{c_2}{\omega_2}\right)\right)+24 \pi  c \cosh (2 \pi  k)+2 \pi  c \cosh (4 \pi  k) \nonumber \\
            & +6 \sinh \left(2 \left(\log \left(\frac{c_2}{\omega_2}\right)+\pi  k\right)\right)+\sinh \left(2 \left(\log \left(\frac{c_2}{\omega_2}\right)+2 \pi  k\right)\right) \nonumber \\
            & -6 \sinh \left(2 \pi  k-2 \log \left(\frac{c_2}{\omega_2}\right)\right)-\sinh \left(4 \pi  k-2 \log \left(\frac{c_2}{\omega_2}\right)\right) \nonumber \\
            & -6 \sinh \left(2 \log \left(\frac{c_2}{\omega_2}\right)\right)-42 \pi  c\Bigg]
  \end{align}
}
\end{itemize}

\section{Plots of slope-to-height ratio of the potential}

\begin{figure}[!h]
    \centering
    \subfigure[]{\includegraphics[width=0.45\textwidth]{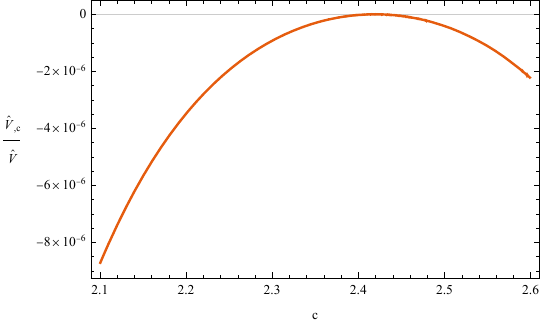}}
    \hfill
    \subfigure[]{\includegraphics[width=0.45\textwidth]{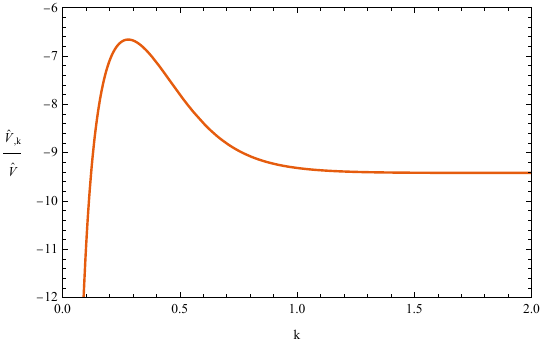}}
    \caption{In the left plot, where only the region of slow roll interest has been shown, all the curves coincide for $k\in[10^{-3},10]$. Same happens in the right plot for $c\in[10^{-3},10]$, with the curves asymptoting to the value of $-9.42$ beyond this range. These plots thus demonstrate that the slope-to-height ratio of $\hat{V}(c,k)$, when evaluated with respect to one modulus, is independent of the value of the other modulus.}
    \label{fig:app1}
\end{figure}

The nature of $(\hat{V}_{,c}/\hat{V})$ and $(\hat{V}_{,k}/\hat{V})$, which appear respectively in \eqref{eq:ceqexact} and \eqref{eq:keqexact}, has been shown graphically in Fig. \ref{fig:app1}.

\section{Physical intuition behind unidirectional slow-roll} \label{sec:appendixC}

In this appendix, we briefly discuss the physical reasoning behind our consideration of slow-roll along either field direction at a time, with the other field value held approximately constant. First, we focus on slow-roll along the $c$-direction as considered in Sec. \ref{subsec:cinf}, for which the fixed nature of $k$ might be motivated in a step-by-step manner as follows.

\begin{enumerate}
        \item In Fig. \ref{fig:dsset}, we have plotted the potential and the non-canonical coefficients in the dS regime. In Fig. \ref{subfig:dSAck}, focusing on the region $c\in(2,4)$, we see that the cross-coefficient $\hat{\tilde{A}}_{ck}(c,k)$ becomes almost flat along both $c$ and $k$ for all values of $k$, thus rendering both $\hat{\tilde{A}}_{ck,c}$ and $\hat{\tilde{A}}_{ck,k}$ negligibly small.
        \item On the other hand, the value of $\hat{\tilde{A}}_{ck}$ is of the typical order $-\mathcal{O}(10)$ in this region.
        \item Next, as \eqref{eq:friedmot1} and \eqref{eq:friedmot2} are second order ODEs, we can freely choose a vanishing initial condition $\dot{k}_{\rm in}\to0$ for the rolling speed.
        \item Implementing the three pieces of information above in \eqref{eq:friedmot1} and \eqref{eq:friedmot2}, the initial stage of the coupled dynamics may be well approximated by
        \begin{equation} \label{eq:appc1}
        2\hat{\tilde{A}}_{cc}(\ddot{c}+3H\dot{c})+\hat{\tilde{A}}_{ck}\ddot{k}-2k'^2\hat{V}_{,c}\approx0\:,
        \end{equation}
        \begin{equation} \label{eq:appc2}
        2\hat{\tilde{A}}_{kk}\ddot{k}+\hat{\tilde{A}}_{ck}(\ddot{c}+3H\dot{c})-2k'^2\hat{V}_{,k}\approx0\:.
        \end{equation}
        \item Now, for $k\sim1.0$ and $c$ starting close to the inflection point $c_{\rm in}\approx2.41994$, the typical magnitudes of the three non-canonical coefficients in the equations above turn out to be $\hat{\tilde{A}}_{cc}\sim10^{-4}$, $\hat{\tilde{A}}_{kk}\sim321$, and $\hat{\tilde{A}}_{ck}\sim-15$. Plugging these values in \eqref{eq:appc2}, we get the order-of-magnitude estimate
        \begin{equation}
        \ddot{c}+3H\dot{c}\sim(15\ddot{k}+2k'^2\hat{V}_{,c})\times10^4\:.
        \end{equation}
        Using this in \eqref{eq:appc1}, we get an order-of-magnitude estimate for $\ddot{k}$ as
        \begin{equation}
        \ddot{k}\sim-k'^2(30\hat{V}_{,c}+2\hat{V}_{,k})\times10^{-6}\:.
        \end{equation}
        \item Referring to Fig. \ref{subfig:dSV}, we see that the slopes of the potential $\hat{V}(c,k)$ along both $c$ and $k$ become very small in the region mentioned above in (e), \textit{i.e.}, both $|\hat{V}_{,c}|$ and $|\hat{V}_{,k}|$ are $\ll\mathcal{O}(1)$. In fact, a direct computation reveals $|\hat{V}_{,c}|\sim\mathcal{O}(10^{-15})$ and $|\hat{V}_{,k}|\sim\mathcal{O}(10^{-8})$. So, overall, $|\ddot{k}|\ll\mathcal{O}(k'^2)$. This, together with the previously chosen initial condition $\dot{k}_{in}\to0$, essentially makes the time-variation of the $k$-field negligibly small, which effectively leads us to the $c$-driven case which is the focus of Sec. \ref{subsec:cinf}. 
    \end{enumerate}
    Finally, while the arguments above have been presented for $k\sim1.0$ for the purpose of demonstration, it is easy to see that any change in the starting value of $k$ affects only the two quantities $\hat{\tilde{A}}_{kk}$ and $\hat{V}_{,k}$. For $k\to0$, both $\hat{A}_{kk}\to-\infty$ and $\hat{V}_{,k}\to-\infty$, as observed in Figs. \ref{subfig:dSAkk} and \ref{subfig:dSV}. But for any positive value of $k$ that is sufficiently away from this divergence (\textit{e.g.} $k\gtrsim0.1$), the order of magnitude calculations presented above remain valid and render $\ddot{k}$ quite small. This, in turn, justifies most of the $k$-values considered in Table \ref{tab:table1}, with the possible exception of the two smallest values. However, we have included those values in our analysis from an inclusive standpoint, in case they too might be motivated physically in the future in extended frameworks.

    In a similar spirit, the fixed nature of $c$ in case of $k$-driven slow-roll may be demonstrated. Within the range $c\sim0.1-1.0$ and $k\sim7.0-11.0$ as chosen in Sec. \ref{subsec:kinf}, choosing $\dot{c}_{\rm in}\to0$ allows us to approximate the initial dynamics described by \eqref{eq:friedmot1} and \eqref{eq:friedmot2} by
    \begin{equation}
        2\hat{\tilde{A}}_{kk}(\ddot{k}+3H\dot{k})+\hat{\tilde{A}}_{ck}\ddot{c}-2k'^2V_{,k}\approx0\:,
    \end{equation}
    \begin{equation}
        2\hat{\tilde{A}}_{cc}\ddot{c}+\hat{\tilde{A}}_{ck}(\ddot{k}+3H\dot{k})-2k'^2V_{,c}\approx0\:.
    \end{equation}
    If, say, $c\sim0.5$ and $k\sim7.0$ (for demonstration), then the typical magnitudes of the quantities involved are $\hat{\tilde{A}}_{kk}\sim325$, $\hat{\tilde{A}}_{cc}\sim-6$, and $\hat{\tilde{A}}_{ck}\sim-60$. Using the first equation to eliminate $(\ddot{k}+3H\dot{k})$ in the second equation, the acceleration of the $c$-field is of the order
    \begin{equation}
        \ddot{c}\sim0.1\times k'^2(\hat{V}_{,c}+0.1\times\hat{V}_{,k})\:.
    \end{equation}
    In this field zone of interest, we obtain $|\hat{V}_{,c}|\sim\mathcal{O}(10^{-34})$ and $|\hat{V}_{,k}|\sim\mathcal{O}(10^{-32})$, which are vanishingly small. Thus, $\ddot{c}$ is negligibly small, which, together with the initial speed $\dot{c}_{\rm in}\to0$, effectively freezes the $c$-field and reduces the scenario to the one considered in Sec. \ref{subsec:kinf}.

\bibliographystyle{JHEP} 
\bibliography{nestwarp_bibliography} 

\providecommand{\href}[2]{#2}\begingroup\raggedright\begin{thebibliography}{10}

\bibitem{Randall:1999ee}
L.~Randall and R.~Sundrum, \emph{{A Large mass hierarchy from a small extra
  dimension}}, \href{https://doi.org/10.1103/PhysRevLett.83.3370}{\emph{Phys.
  Rev. Lett.} {\bfseries 83} (1999) 3370}
  [\href{https://arxiv.org/abs/hep-ph/9905221}{{\ttfamily hep-ph/9905221}}].

\bibitem{Goldberger:1999uk}
W.D.~Goldberger and M.B.~Wise, \emph{{Modulus stabilization with bulk fields}},
  \href{https://doi.org/10.1103/PhysRevLett.83.4922}{\emph{Phys. Rev. Lett.}
  {\bfseries 83} (1999) 4922}
  [\href{https://arxiv.org/abs/hep-ph/9907447}{{\ttfamily hep-ph/9907447}}].

\bibitem{Brevik:2000vt}
I.H.~Brevik, K.A.~Milton, S.~Nojiri and S.D.~Odintsov, \emph{{Quantum
  (in)stability of a brane world AdS(5) universe at nonzero temperature}},
  \href{https://doi.org/10.1016/S0550-3213(01)00026-8}{\emph{Nucl. Phys. B}
  {\bfseries 599} (2001) 305}
  [\href{https://arxiv.org/abs/hep-th/0010205}{{\ttfamily hep-th/0010205}}].

\bibitem{Dey:2006px}
A.~Dey, D.~Maity and S.~SenGupta, \emph{{A Critical Analysis of Goldberger-Wise
  Stabilization in Randall-Sundrum Scenario}},
  \href{https://doi.org/10.1103/PhysRevD.75.107901}{\emph{Phys. Rev. D}
  {\bfseries 75} (2007) 107901}
  [\href{https://arxiv.org/abs/hep-th/0611262}{{\ttfamily hep-th/0611262}}].

\bibitem{Chakraborty:2016gpg}
S.~Chakraborty and S.~SenGupta, \emph{{Gravity stabilizes itself}},
  \href{https://doi.org/10.1140/epjc/s10052-017-5138-5}{\emph{Eur. Phys. J. C}
  {\bfseries 77} (2017) 573}
  [\href{https://arxiv.org/abs/1701.01032}{{\ttfamily 1701.01032}}].

\bibitem{Das:2017htt}
A.~Das, H.~Mukherjee, T.~Paul and S.~SenGupta, \emph{{Radion stabilization in
  higher curvature warped spacetime}},
  \href{https://doi.org/10.1140/epjc/s10052-018-5603-9}{\emph{Eur. Phys. J. C}
  {\bfseries 78} (2018) 108}
  [\href{https://arxiv.org/abs/1701.01571}{{\ttfamily 1701.01571}}].

\bibitem{Elahi:2022hpj}
S.G.~Elahi, S.S.~Mandal and S.~SenGupta, \emph{{Novel modulus stabilization
  mechanism in higher dimensional f(R) gravity}},
  \href{https://doi.org/10.1103/PhysRevD.108.044062}{\emph{Phys. Rev. D}
  {\bfseries 108} (2023) 044062}
  [\href{https://arxiv.org/abs/2212.03276}{{\ttfamily 2212.03276}}].

\bibitem{Goldberger:1999un}
W.D.~Goldberger and M.B.~Wise, \emph{{Phenomenology of a stabilized modulus}},
  \href{https://doi.org/10.1016/S0370-2693(00)00099-X}{\emph{Phys. Lett. B}
  {\bfseries 475} (2000) 275}
  [\href{https://arxiv.org/abs/hep-ph/9911457}{{\ttfamily hep-ph/9911457}}].

\bibitem{Davoudiasl:1999tf}
H.~Davoudiasl, J.L.~Hewett and T.G.~Rizzo, \emph{{Bulk gauge fields in the
  Randall-Sundrum model}},
  \href{https://doi.org/10.1016/S0370-2693(99)01430-6}{\emph{Phys. Lett. B}
  {\bfseries 473} (2000) 43}
  [\href{https://arxiv.org/abs/hep-ph/9911262}{{\ttfamily hep-ph/9911262}}].

\bibitem{Davoudiasl:1999jd}
H.~Davoudiasl, J.L.~Hewett and T.G.~Rizzo, \emph{{Phenomenology of the
  Randall-Sundrum Gauge Hierarchy Model}},
  \href{https://doi.org/10.1103/PhysRevLett.84.2080}{\emph{Phys. Rev. Lett.}
  {\bfseries 84} (2000) 2080}
  [\href{https://arxiv.org/abs/hep-ph/9909255}{{\ttfamily hep-ph/9909255}}].

\bibitem{DeWolfe:1999cp}
O.~DeWolfe, D.Z.~Freedman, S.S.~Gubser and A.~Karch, \emph{{Modeling the
  fifth-dimension with scalars and gravity}},
  \href{https://doi.org/10.1103/PhysRevD.62.046008}{\emph{Phys. Rev. D}
  {\bfseries 62} (2000) 046008}
  [\href{https://arxiv.org/abs/hep-th/9909134}{{\ttfamily hep-th/9909134}}].

\bibitem{Chang:1999nh}
S.~Chang, J.~Hisano, H.~Nakano, N.~Okada and M.~Yamaguchi, \emph{{Bulk standard
  model in the Randall-Sundrum background}},
  \href{https://doi.org/10.1103/PhysRevD.62.084025}{\emph{Phys. Rev. D}
  {\bfseries 62} (2000) 084025}
  [\href{https://arxiv.org/abs/hep-ph/9912498}{{\ttfamily hep-ph/9912498}}].

\bibitem{Csaki:2000zn}
C.~Csaki, M.L.~Graesser and G.D.~Kribs, \emph{{Radion dynamics and electroweak
  physics}}, \href{https://doi.org/10.1103/PhysRevD.63.065002}{\emph{Phys. Rev.
  D} {\bfseries 63} (2001) 065002}
  [\href{https://arxiv.org/abs/hep-th/0008151}{{\ttfamily hep-th/0008151}}].

\bibitem{Huber:2000ie}
S.J.~Huber and Q.~Shafi, \emph{{Fermion masses, mixings and proton decay in a
  Randall-Sundrum model}},
  \href{https://doi.org/10.1016/S0370-2693(00)01399-X}{\emph{Phys. Lett. B}
  {\bfseries 498} (2001) 256}
  [\href{https://arxiv.org/abs/hep-ph/0010195}{{\ttfamily hep-ph/0010195}}].

\bibitem{Agashe:2003zs}
K.~Agashe, A.~Delgado, M.J.~May and R.~Sundrum, \emph{{RS1, custodial isospin
  and precision tests}},
  \href{https://doi.org/10.1088/1126-6708/2003/08/050}{\emph{JHEP} {\bfseries
  08} (2003) 050} [\href{https://arxiv.org/abs/hep-ph/0308036}{{\ttfamily
  hep-ph/0308036}}].

\bibitem{Luty:2004ye}
M.A.~Luty and T.~Okui, \emph{{Conformal technicolor}},
  \href{https://doi.org/10.1088/1126-6708/2006/09/070}{\emph{JHEP} {\bfseries
  09} (2006) 070} [\href{https://arxiv.org/abs/hep-ph/0409274}{{\ttfamily
  hep-ph/0409274}}].

\bibitem{Davoudiasl:2005uu}
H.~Davoudiasl, B.~Lillie and T.G.~Rizzo, \emph{{Off-the-wall Higgs in the
  universal Randall-Sundrum model}},
  \href{https://doi.org/10.1088/1126-6708/2006/08/042}{\emph{JHEP} {\bfseries
  08} (2006) 042} [\href{https://arxiv.org/abs/hep-ph/0508279}{{\ttfamily
  hep-ph/0508279}}].

\bibitem{Chacko:2013dra}
Z.~Chacko, R.K.~Mishra and D.~Stolarski, \emph{{Dynamics of a Stabilized Radion
  and Duality}}, \href{https://doi.org/10.1007/JHEP09(2013)121}{\emph{JHEP}
  {\bfseries 09} (2013) 121} [\href{https://arxiv.org/abs/1304.1795}{{\ttfamily
  1304.1795}}].

\bibitem{Ahmed:2019zxm}
A.~Ahmed, A.~Carmona, J.~Castellano~Ruiz, Y.~Chung and M.~Neubert,
  \emph{{Dynamical origin of fermion bulk masses in a warped extra dimension}},
  \href{https://doi.org/10.1007/JHEP08(2019)045}{\emph{JHEP} {\bfseries 08}
  (2019) 045} [\href{https://arxiv.org/abs/1905.09833}{{\ttfamily
  1905.09833}}].

\bibitem{Lee:2021wau}
S.J.~Lee, Y.~Nakai and M.~Suzuki, \emph{{Multiple hierarchies from a warped
  extra dimension}}, \href{https://doi.org/10.1007/JHEP02(2022)050}{\emph{JHEP}
  {\bfseries 02} (2022) 050}
  [\href{https://arxiv.org/abs/2109.10938}{{\ttfamily 2109.10938}}].

\bibitem{Frank:2023lxf}
M.~Frank, N.~Pourtolami and M.~Toharia, \emph{{A nearly Dirichlet Higgs for
  lower-scale warped extra dimensions}},
  \href{https://doi.org/10.1016/j.physletb.2023.138084}{\emph{Phys. Lett. B}
  {\bfseries 844} (2023) 138084}
  [\href{https://arxiv.org/abs/2305.09567}{{\ttfamily 2305.09567}}].

\bibitem{ATLAS:2011ab}
{\scshape ATLAS} collaboration, \emph{{Search for extra dimensions using
  diphoton events in 7 TeV proton\textendash{}proton collisions with the ATLAS
  detector}}, \href{https://doi.org/10.1016/j.physletb.2012.03.022}{\emph{Phys.
  Lett. B} {\bfseries 710} (2012) 538}
  [\href{https://arxiv.org/abs/1112.2194}{{\ttfamily 1112.2194}}].

\bibitem{ATLAS:2012hvw}
{\scshape ATLAS} collaboration, \emph{{Search for Extra Dimensions in diphoton
  events using proton-proton collisions recorded at $\sqrt{s}=7$ TeV with the
  ATLAS detector at the LHC}},
  \href{https://doi.org/10.1088/1367-2630/15/4/043007}{\emph{New J. Phys.}
  {\bfseries 15} (2013) 043007}
  [\href{https://arxiv.org/abs/1210.8389}{{\ttfamily 1210.8389}}].

\bibitem{ATLAS:2014pcp}
{\scshape ATLAS} collaboration, \emph{{Search for high-mass dilepton resonances
  in pp collisions at $\sqrt{s}=8$ TeV with the ATLAS detector}},
  \href{https://doi.org/10.1103/PhysRevD.90.052005}{\emph{Phys. Rev. D}
  {\bfseries 90} (2014) 052005}
  [\href{https://arxiv.org/abs/1405.4123}{{\ttfamily 1405.4123}}].

\bibitem{CMS:2014mws}
{\scshape CMS} collaboration, \emph{{Search for massive resonances decaying
  into pairs of boosted bosons in semi-leptonic final states at $\sqrt{s} =$ 8
  TeV}}, \href{https://doi.org/10.1007/JHEP08(2014)174}{\emph{JHEP} {\bfseries
  08} (2014) 174} [\href{https://arxiv.org/abs/1405.3447}{{\ttfamily
  1405.3447}}].

\bibitem{ATLAS:2015shg}
{\scshape ATLAS} collaboration, \emph{{Search for high-mass diphoton resonances
  in $pp$ collisions at $\sqrt{s}=8$ TeV with the ATLAS detector}},
  \href{https://doi.org/10.1103/PhysRevD.92.032004}{\emph{Phys. Rev. D}
  {\bfseries 92} (2015) 032004}
  [\href{https://arxiv.org/abs/1504.05511}{{\ttfamily 1504.05511}}].

\bibitem{CMS:2015cwa}
{\scshape CMS} collaboration, \emph{{Search for High-Mass Diphoton Resonances
  in pp Collisions at sqrt(s)=8 TeV with the CMS Detector}}, .

\bibitem{CMS:2016crm}
{\scshape CMS} collaboration, \emph{{Search for resonant production of high
  mass photon pairs using $12.9\,\mathrm{fb^{-1}}$ of proton-proton collisions
  at $\sqrt{s} = 13~\mathrm{TeV}$ and combined interpretation of searches at 8
  and 13 TeV}}, .

\bibitem{ATLAS:2017zuf}
{\scshape ATLAS} collaboration, \emph{{Search for diboson resonances with
  boson-tagged jets in $pp$ collisions at $\sqrt{s}=13$ TeV with the ATLAS
  detector}}, \href{https://doi.org/10.1016/j.physletb.2017.12.011}{\emph{Phys.
  Lett. B} {\bfseries 777} (2018) 91}
  [\href{https://arxiv.org/abs/1708.04445}{{\ttfamily 1708.04445}}].

\bibitem{ATLAS:2018rnh}
{\scshape ATLAS} collaboration, \emph{{Search for pair production of Higgs
  bosons in the $b\bar{b}b\bar{b}$ final state using proton-proton collisions
  at $\sqrt{s} = 13$ TeV with the ATLAS detector}},
  \href{https://doi.org/10.1007/JHEP01(2019)030}{\emph{JHEP} {\bfseries 01}
  (2019) 030} [\href{https://arxiv.org/abs/1804.06174}{{\ttfamily
  1804.06174}}].

\bibitem{ATLAS:2019erb}
{\scshape ATLAS} collaboration, \emph{{Search for high-mass dilepton resonances
  using 139 fb$^{-1}$ of $pp$ collision data collected at $\sqrt{s}=$13 TeV
  with the ATLAS detector}},
  \href{https://doi.org/10.1016/j.physletb.2019.07.016}{\emph{Phys. Lett. B}
  {\bfseries 796} (2019) 68}
  [\href{https://arxiv.org/abs/1903.06248}{{\ttfamily 1903.06248}}].

\bibitem{Arun:2014dga}
M.T.~Arun, D.~Choudhury, A.~Das and S.~SenGupta, \emph{{Graviton modes in
  multiply warped geometry}},
  \href{https://doi.org/10.1016/j.physletb.2015.05.008}{\emph{Phys. Lett. B}
  {\bfseries 746} (2015) 266}
  [\href{https://arxiv.org/abs/1410.5591}{{\ttfamily 1410.5591}}].

\bibitem{Choudhury:2006nj}
D.~Choudhury and S.~SenGupta, \emph{{Living on the edge in a spacetime with
  multiple warping}},
  \href{https://doi.org/10.1103/PhysRevD.76.064030}{\emph{Phys. Rev. D}
  {\bfseries 76} (2007) 064030}
  [\href{https://arxiv.org/abs/hep-th/0612246}{{\ttfamily hep-th/0612246}}].

\bibitem{Randjbar-Daemi:2000bjr}
S.~Randjbar-Daemi and M.E.~Shaposhnikov, \emph{{On some new warped brane world
  solutions in higher dimensions}},
  \href{https://doi.org/10.1016/S0370-2693(00)01060-1}{\emph{Phys. Lett. B}
  {\bfseries 491} (2000) 329}
  [\href{https://arxiv.org/abs/hep-th/0008087}{{\ttfamily hep-th/0008087}}].

\bibitem{PhysRevD.64.044021}
P.~Kanti, R.~Madden and K.A.~Olive, \emph{6-dimensional brane world model},
  \href{https://doi.org/10.1103/PhysRevD.64.044021}{\emph{Phys. Rev. D}
  {\bfseries 64} (2001) 044021}.

\bibitem{PhysRevLett.90.101601}
T.~Gherghetta and A.~Kehagias, \emph{Anomaly-free brane worlds in seven
  dimensions}, \href{https://doi.org/10.1103/PhysRevLett.90.101601}{\emph{Phys.
  Rev. Lett.} {\bfseries 90} (2003) 101601}.

\bibitem{Kaloper:2004cy}
N.~Kaloper, \emph{{Origami world}},
  \href{https://doi.org/10.1088/1126-6708/2004/05/061}{\emph{JHEP} {\bfseries
  05} (2004) 061} [\href{https://arxiv.org/abs/hep-th/0403208}{{\ttfamily
  hep-th/0403208}}].

\bibitem{PhysRevD.72.064008}
B.~Cuadros-Melgar and E.~Papantonopoulos, \emph{Need of dark energy for
  dynamical compactification of extra dimensions on the brane},
  \href{https://doi.org/10.1103/PhysRevD.72.064008}{\emph{Phys. Rev. D}
  {\bfseries 72} (2005) 064008}.

\bibitem{PhysRevD.77.124046}
K.L.~McDonald, \emph{Little randall-sundrum model and a multiply warped
  spacetime}, \href{https://doi.org/10.1103/PhysRevD.77.124046}{\emph{Phys.
  Rev. D} {\bfseries 77} (2008) 124046}.

\bibitem{Archer:2010bm}
P.R.~Archer and S.J.~Huber, \emph{{Reducing Constraints in a Higher Dimensional
  Extension of the Randall and Sundrum Model}},
  \href{https://doi.org/10.1007/JHEP03(2011)018}{\emph{JHEP} {\bfseries 03}
  (2011) 018} [\href{https://arxiv.org/abs/1010.3588}{{\ttfamily 1010.3588}}].

\bibitem{Feng:2015sfa}
S.-F.~Feng, C.-Y.~Huang, Y.-C.~Huang, X.~Liu and Y.-J.~Zhao, \emph{{Symmetry of
  Generalized Randall-Sundrum Model and Distribution of 3-Branes in
  Six-Dimensional Spacetime}},
  \href{https://arxiv.org/abs/1506.03598}{{\ttfamily 1506.03598}}.

\bibitem{Meiers:2017ltj}
M.~Meiers, L.~Bovard and R.~Mann, \emph{{Charged Randall\textendash{}Sundrum
  black holes in higher dimensions}},
  \href{https://doi.org/10.1088/1361-6382/aa9770}{\emph{Class. Quant. Grav.}
  {\bfseries 35} (2018) 025006}
  [\href{https://arxiv.org/abs/1708.01603}{{\ttfamily 1708.01603}}].

\bibitem{Wan:2020smy}
J.-J.~Wan, Z.-Q.~Cui, W.-B.~Feng and Y.-X.~Liu, \emph{{Smooth braneworld in
  $6$-dimensional asymptotically AdS spacetime}},
  \href{https://doi.org/10.1007/JHEP05(2021)017}{\emph{JHEP} {\bfseries 05}
  (2021) 017} [\href{https://arxiv.org/abs/2010.05016}{{\ttfamily
  2010.05016}}].

\bibitem{Arun:2015ubr}
M.T.~Arun and P.~Saha, \emph{{Gravitons in multiply warped scenarios: At 750
  GeV and beyond}},
  \href{https://doi.org/10.1007/s12043-017-1387-y}{\emph{Pramana} {\bfseries
  88} (2017) 93} [\href{https://arxiv.org/abs/1512.06335}{{\ttfamily
  1512.06335}}].

\bibitem{Hundi:2011dc}
R.S.~Hundi and S.~SenGupta, \emph{{Fermion mass hierarchy in a multiple warped
  braneworld model}},
  \href{https://doi.org/10.1088/0954-3899/40/7/075002}{\emph{J. Phys. G}
  {\bfseries 40} (2013) 075002}
  [\href{https://arxiv.org/abs/1111.1106}{{\ttfamily 1111.1106}}].

\bibitem{Arun:2016csq}
M.T.~Arun and D.~Choudhury, \emph{{Stabilization of moduli in spacetime with
  nested warping and the UED}},
  \href{https://doi.org/10.1016/j.nuclphysb.2017.08.004}{\emph{Nucl. Phys. B}
  {\bfseries 923} (2017) 258}
  [\href{https://arxiv.org/abs/1606.00642}{{\ttfamily 1606.00642}}].

\bibitem{Bhaumik:2022xtd}
A.~Bhaumik and S.~SenGupta, \emph{{Moduli stabilization with bulk scalar in
  nested doubly warped braneworld model}},
  \href{https://doi.org/10.1140/epjc/s10052-022-10973-y}{\emph{Eur. Phys. J. C}
  {\bfseries 82} (2022) 1079}
  [\href{https://arxiv.org/abs/2201.10503}{{\ttfamily 2201.10503}}].

\bibitem{Das:2011fb}
A.~Das, R.S.~Hundi and S.~SenGupta, \emph{{Bulk Higgs and Gauge fields in a
  multiply warped braneworld model}},
  \href{https://doi.org/10.1103/PhysRevD.83.116003}{\emph{Phys. Rev. D}
  {\bfseries 83} (2011) 116003}
  [\href{https://arxiv.org/abs/1105.1064}{{\ttfamily 1105.1064}}].

\bibitem{Chakraborty:2014xda}
S.~Chakraborty and S.~SenGupta, \emph{{Bulk scalar field in warped extra
  dimensional models}},
  \href{https://doi.org/10.1103/PhysRevD.89.126001}{\emph{Phys. Rev. D}
  {\bfseries 89} (2014) 126001}
  [\href{https://arxiv.org/abs/1401.3279}{{\ttfamily 1401.3279}}].

\bibitem{Arun:2015kva}
M.T.~Arun and D.~Choudhury, \emph{{Bulk gauge and matter fields in nested
  warping: I. the formalism}},
  \href{https://doi.org/10.1007/JHEP09(2015)202}{\emph{JHEP} {\bfseries 09}
  (2015) 202} [\href{https://arxiv.org/abs/1501.06118}{{\ttfamily
  1501.06118}}].

\bibitem{Arun:2016ela}
M.T.~Arun and D.~Choudhury, \emph{{Bulk gauge and matter fields in nested
  warping: II. Symmetry Breaking and phenomenological consequences}},
  \href{https://doi.org/10.1007/JHEP04(2016)133}{\emph{JHEP} {\bfseries 04}
  (2016) 133} [\href{https://arxiv.org/abs/1601.02321}{{\ttfamily
  1601.02321}}].

\bibitem{Barman:2022qix}
B.~Barman, A.~Das and S.~Sengupta, \emph{{New $W$-Boson mass in the light of
  doubly warped braneworld model}},
  \href{https://arxiv.org/abs/2205.01699}{{\ttfamily 2205.01699}}.

\bibitem{Shiromizu:1999wj}
T.~Shiromizu, K.-i.~Maeda and M.~Sasaki, \emph{{The Einstein equation on the
  3-brane world}},
  \href{https://doi.org/10.1103/PhysRevD.62.024012}{\emph{Phys. Rev. D}
  {\bfseries 62} (2000) 024012}
  [\href{https://arxiv.org/abs/gr-qc/9910076}{{\ttfamily gr-qc/9910076}}].

\bibitem{Ida:2001qw}
D.~Ida, T.~Shiromizu and H.~Ochiai, \emph{{Semiclassical instability of the
  brane world: Randall-Sundrum bubbles}},
  \href{https://doi.org/10.1103/PhysRevD.65.023504}{\emph{Phys. Rev. D}
  {\bfseries 65} (2002) 023504}
  [\href{https://arxiv.org/abs/hep-th/0108056}{{\ttfamily hep-th/0108056}}].

\bibitem{Charmousis:2003sq}
C.~Charmousis and J.-F.~Dufaux, \emph{{Gauss-Bonnet gravity renders negative
  tension brane worlds unstable}},
  \href{https://doi.org/10.1103/PhysRevD.70.106002}{\emph{Phys. Rev. D}
  {\bfseries 70} (2004) 106002}
  [\href{https://arxiv.org/abs/hep-th/0311267}{{\ttfamily hep-th/0311267}}].

\bibitem{Das:2007qn}
S.~Das, D.~Maity and S.~SenGupta, \emph{{Cosmological constant, brane tension
  and large hierarchy in a generalized Randall-Sundrum braneworld scenario}},
  \href{https://doi.org/10.1088/1126-6708/2008/05/042}{\emph{JHEP} {\bfseries
  05} (2008) 042} [\href{https://arxiv.org/abs/0711.1744}{{\ttfamily
  0711.1744}}].

\bibitem{Koley:2008hs}
R.~Koley, J.~Mitra and S.~SenGupta, \emph{{Modulus stabilization of generalized
  Randall Sundrum model with bulk scalar field}},
  \href{https://doi.org/10.1209/0295-5075/85/41001}{\emph{EPL} {\bfseries 85}
  (2009) 41001} [\href{https://arxiv.org/abs/0809.4102}{{\ttfamily
  0809.4102}}].

\bibitem{Banerjee:2017jyk}
I.~Banerjee and S.~SenGupta, \emph{{Modulus stabilization in a non-flat warped
  braneworld scenario}},
  \href{https://doi.org/10.1140/epjc/s10052-017-4857-y}{\emph{Eur. Phys. J. C}
  {\bfseries 77} (2017) 277}
  [\href{https://arxiv.org/abs/1705.05015}{{\ttfamily 1705.05015}}].

\bibitem{Banerjee:2018kcz}
I.~Banerjee, S.~Chakraborty and S.~SenGupta, \emph{{Radion induced inflation on
  nonflat brane and modulus stabilization}},
  \href{https://doi.org/10.1103/PhysRevD.99.023515}{\emph{Phys. Rev. D}
  {\bfseries 99} (2019) 023515}
  [\href{https://arxiv.org/abs/1806.11327}{{\ttfamily 1806.11327}}].

\bibitem{Banerjee:2020uil}
I.~Banerjee, T.~Paul and S.~SenGupta, \emph{{Bouncing cosmology in a curved
  braneworld}},
  \href{https://doi.org/10.1088/1475-7516/2021/02/041}{\emph{JCAP} {\bfseries
  02} (2021) 041} [\href{https://arxiv.org/abs/2011.11886}{{\ttfamily
  2011.11886}}].

\bibitem{Bhaumik:2023tmg}
A.~Bhaumik and S.~SenGupta, \emph{{Nested warped geometry in a non-flat
  braneworld scenario}},
  \href{https://doi.org/10.1140/epjc/s10052-023-11795-2}{\emph{Eur. Phys. J. C}
  {\bfseries 83} (2023) 788}
  [\href{https://arxiv.org/abs/2301.00698}{{\ttfamily 2301.00698}}].

\bibitem{Randall:1999vf}
L.~Randall and R.~Sundrum, \emph{{An Alternative to compactification}},
  \href{https://doi.org/10.1103/PhysRevLett.83.4690}{\emph{Phys. Rev. Lett.}
  {\bfseries 83} (1999) 4690}
  [\href{https://arxiv.org/abs/hep-th/9906064}{{\ttfamily hep-th/9906064}}].

\bibitem{xact}
J.M.~Martin-Garcia et~al., \emph{``{xAct: Efficient tensor computation algebra
  for Mathematica}.''} \url{http://xact.es/}, 2002-2024.

\bibitem{Garriga:1999hf}
J.~Garriga, V.F.~Mukhanov, K.D.~Olum and A.~Vilenkin, \emph{{Eternal inflation,
  black holes, and the future of civilizations}},
  \href{https://doi.org/10.1023/A:1003602000709}{\emph{Int. J. Theor. Phys.}
  {\bfseries 39} (2000) 1887}
  [\href{https://arxiv.org/abs/astro-ph/9909143}{{\ttfamily
  astro-ph/9909143}}].

\bibitem{Chung:2003iu}
D.J.H.~Chung, G.~Shiu and M.~Trodden, \emph{{Running of the scalar spectral
  index from inflationary models}},
  \href{https://doi.org/10.1103/PhysRevD.68.063501}{\emph{Phys. Rev. D}
  {\bfseries 68} (2003) 063501}
  [\href{https://arxiv.org/abs/astro-ph/0305193}{{\ttfamily
  astro-ph/0305193}}].

\bibitem{Liddle:1993fq}
A.R.~Liddle and D.H.~Lyth, \emph{{The Cold dark matter density perturbation}},
  \href{https://doi.org/10.1016/0370-1573(93)90114-S}{\emph{Phys. Rept.}
  {\bfseries 231} (1993) 1}
  [\href{https://arxiv.org/abs/astro-ph/9303019}{{\ttfamily
  astro-ph/9303019}}].

\bibitem{Liddle:2003as}
A.R.~Liddle and S.M.~Leach, \emph{{How long before the end of inflation were
  observable perturbations produced?}},
  \href{https://doi.org/10.1103/PhysRevD.68.103503}{\emph{Phys. Rev. D}
  {\bfseries 68} (2003) 103503}
  [\href{https://arxiv.org/abs/astro-ph/0305263}{{\ttfamily
  astro-ph/0305263}}].

\bibitem{Leach:2002dw}
S.M.~Leach and A.R.~Liddle, \emph{{Microwave background constraints on
  inflationary parameters}},
  \href{https://doi.org/10.1046/j.1365-8711.2003.06445.x}{\emph{Mon. Not. Roy.
  Astron. Soc.} {\bfseries 341} (2003) 1151}
  [\href{https://arxiv.org/abs/astro-ph/0207213}{{\ttfamily
  astro-ph/0207213}}].

\bibitem{Planck:2018nkj}
{\scshape Planck} collaboration, \emph{{Planck 2018 results. I. Overview and
  the cosmological legacy of Planck}},
  \href{https://doi.org/10.1051/0004-6361/201833880}{\emph{Astron. Astrophys.}
  {\bfseries 641} (2020) A1}
  [\href{https://arxiv.org/abs/1807.06205}{{\ttfamily 1807.06205}}].

\bibitem{Planck:2018vyg}
{\scshape Planck} collaboration, \emph{{Planck 2018 results. VI. Cosmological
  parameters}},
  \href{https://doi.org/10.1051/0004-6361/201833910}{\emph{Astron. Astrophys.}
  {\bfseries 641} (2020) A6}
  [\href{https://arxiv.org/abs/1807.06209}{{\ttfamily 1807.06209}}].

\bibitem{Planck:2018jri}
{\scshape Planck} collaboration, \emph{{Planck 2018 results. X. Constraints on
  inflation}}, \href{https://doi.org/10.1051/0004-6361/201833887}{\emph{Astron.
  Astrophys.} {\bfseries 641} (2020) A10}
  [\href{https://arxiv.org/abs/1807.06211}{{\ttfamily 1807.06211}}].

\bibitem{BICEP:2021xfz}
{\scshape BICEP, Keck} collaboration, \emph{{Improved Constraints on Primordial
  Gravitational Waves using Planck, WMAP, and BICEP/Keck Observations through
  the 2018 Observing Season}},
  \href{https://doi.org/10.1103/PhysRevLett.127.151301}{\emph{Phys. Rev. Lett.}
  {\bfseries 127} (2021) 151301}
  [\href{https://arxiv.org/abs/2110.00483}{{\ttfamily 2110.00483}}].

\bibitem{Tristram:2021tvh}
M.~Tristram et~al., \emph{{Improved limits on the tensor-to-scalar ratio using
  BICEP and Planck data}},
  \href{https://doi.org/10.1103/PhysRevD.105.083524}{\emph{Phys. Rev. D}
  {\bfseries 105} (2022) 083524}
  [\href{https://arxiv.org/abs/2112.07961}{{\ttfamily 2112.07961}}].

\bibitem{Campeti:2022vom}
P.~Campeti and E.~Komatsu, \emph{{New Constraint on the Tensor-to-scalar Ratio
  from the Planck and BICEP/Keck Array Data Using the Profile Likelihood}},
  \href{https://doi.org/10.3847/1538-4357/ac9ea3}{\emph{Astrophys. J.}
  {\bfseries 941} (2022) 110}
  [\href{https://arxiv.org/abs/2205.05617}{{\ttfamily 2205.05617}}].

\bibitem{Lyth:1996im}
D.H.~Lyth, \emph{{What would we learn by detecting a gravitational wave signal
  in the cosmic microwave background anisotropy?}},
  \href{https://doi.org/10.1103/PhysRevLett.78.1861}{\emph{Phys. Rev. Lett.}
  {\bfseries 78} (1997) 1861}
  [\href{https://arxiv.org/abs/hep-ph/9606387}{{\ttfamily hep-ph/9606387}}].

\bibitem{RevModPhys.69.373}
J.E.~Lidsey, A.R.~Liddle, E.W.~Kolb, E.J.~Copeland, T.~Barreiro and M.~Abney,
  \emph{Reconstructing the inflaton potential---an overview},
  \href{https://doi.org/10.1103/RevModPhys.69.373}{\emph{Rev. Mod. Phys.}
  {\bfseries 69} (1997) 373}.

\bibitem{Efstathiou:2005tq}
G.~Efstathiou and K.J.~Mack, \emph{{The Lyth bound revisited}},
  \href{https://doi.org/10.1088/1475-7516/2005/05/008}{\emph{JCAP} {\bfseries
  05} (2005) 008} [\href{https://arxiv.org/abs/astro-ph/0503360}{{\ttfamily
  astro-ph/0503360}}].

\bibitem{Easther:2006qu}
R.~Easther, W.H.~Kinney and B.A.~Powell, \emph{{The Lyth bound and the end of
  inflation}}, \href{https://doi.org/10.1088/1475-7516/2006/08/004}{\emph{JCAP}
  {\bfseries 08} (2006) 004}
  [\href{https://arxiv.org/abs/astro-ph/0601276}{{\ttfamily
  astro-ph/0601276}}].

\bibitem{Anber:2009ua}
M.M.~Anber and L.~Sorbo, \emph{{Naturally inflating on steep potentials through
  electromagnetic dissipation}},
  \href{https://doi.org/10.1103/PhysRevD.81.043534}{\emph{Phys. Rev. D}
  {\bfseries 81} (2010) 043534}
  [\href{https://arxiv.org/abs/0908.4089}{{\ttfamily 0908.4089}}].

\bibitem{Adshead:2012kp}
P.~Adshead and M.~Wyman, \emph{{Chromo-Natural Inflation: Natural inflation on
  a steep potential with classical non-Abelian gauge fields}},
  \href{https://doi.org/10.1103/PhysRevLett.108.261302}{\emph{Phys. Rev. Lett.}
  {\bfseries 108} (2012) 261302}
  [\href{https://arxiv.org/abs/1202.2366}{{\ttfamily 1202.2366}}].

\bibitem{Rezazadeh:2015dia}
K.~Rezazadeh, K.~Karami and S.~Hashemi, \emph{{Tachyon inflation with steep
  potentials}}, \href{https://doi.org/10.1103/PhysRevD.95.103506}{\emph{Phys.
  Rev. D} {\bfseries 95} (2017) 103506}
  [\href{https://arxiv.org/abs/1508.04760}{{\ttfamily 1508.04760}}].

\bibitem{Adshead:2016iix}
P.~Adshead, D.~Blas, C.P.~Burgess, P.~Hayman and S.P.~Patil, \emph{{Magnon
  Inflation: Slow Roll with Steep Potentials}},
  \href{https://doi.org/10.1088/1475-7516/2016/11/009}{\emph{JCAP} {\bfseries
  11} (2016) 009} [\href{https://arxiv.org/abs/1604.06048}{{\ttfamily
  1604.06048}}].

\bibitem{Enqvist:2012qc}
K.~Enqvist, \emph{{Cosmological inflation}},  in \emph{{2010 European School of
  High Energy Physics}}, 1, 2012
  [\href{https://arxiv.org/abs/1201.6164}{{\ttfamily 1201.6164}}].

\bibitem{Verlinde:1999fy}
H.L.~Verlinde, \emph{{Holography and compactification}},
  \href{https://doi.org/10.1016/S0550-3213(00)00224-8}{\emph{Nucl. Phys. B}
  {\bfseries 580} (2000) 264}
  [\href{https://arxiv.org/abs/hep-th/9906182}{{\ttfamily hep-th/9906182}}].

\bibitem{Chan:2000ms}
C.S.~Chan, P.L.~Paul and H.L.~Verlinde, \emph{{A Note on warped string
  compactification}},
  \href{https://doi.org/10.1016/S0550-3213(00)00267-4}{\emph{Nucl. Phys. B}
  {\bfseries 581} (2000) 156}
  [\href{https://arxiv.org/abs/hep-th/0003236}{{\ttfamily hep-th/0003236}}].

\bibitem{Kachru:2002he}
S.~Kachru, M.B.~Schulz and S.~Trivedi, \emph{{Moduli stabilization from fluxes
  in a simple IIB orientifold}},
  \href{https://doi.org/10.1088/1126-6708/2003/10/007}{\emph{JHEP} {\bfseries
  10} (2003) 007} [\href{https://arxiv.org/abs/hep-th/0201028}{{\ttfamily
  hep-th/0201028}}].

\bibitem{Brummer:2005sh}
F.~Brummer, A.~Hebecker and E.~Trincherini, \emph{{The Throat as a
  Randall-Sundrum model with Goldberger-Wise stabilization}},
  \href{https://doi.org/10.1016/j.nuclphysb.2006.01.011}{\emph{Nucl. Phys. B}
  {\bfseries 738} (2006) 283}
  [\href{https://arxiv.org/abs/hep-th/0510113}{{\ttfamily hep-th/0510113}}].

\bibitem{Piao:2004tq}
Y.-S.~Piao and Y.-Z.~Zhang, \emph{{Phantom inflation and primordial
  perturbation spectrum}},
  \href{https://doi.org/10.1103/PhysRevD.70.063513}{\emph{Phys. Rev. D}
  {\bfseries 70} (2004) 063513}
  [\href{https://arxiv.org/abs/astro-ph/0401231}{{\ttfamily
  astro-ph/0401231}}].

\bibitem{Liu:2012iba}
Z.-G.~Liu and Y.-S.~Piao, \emph{{Phantom Inflation in Little Rip}},
  \href{https://doi.org/10.1016/j.physletb.2012.05.027}{\emph{Phys. Lett. B}
  {\bfseries 713} (2012) 53} [\href{https://arxiv.org/abs/1203.4901}{{\ttfamily
  1203.4901}}].

\bibitem{Richarte:2016qqm}
M.G.~Richarte and G.M.~Kremer, \emph{{Cosmological perturbation in the
  transient phantom inflation scenario}},
  \href{https://doi.org/10.1140/epjc/s10052-017-4629-8}{\emph{Eur. Phys. J. C}
  {\bfseries 77} (2017) 51} [\href{https://arxiv.org/abs/1612.03822}{{\ttfamily
  1612.03822}}].

\bibitem{Arkani-Hamed:2003juy}
N.~Arkani-Hamed, P.~Creminelli, S.~Mukohyama and M.~Zaldarriaga, \emph{{Ghost
  inflation}}, \href{https://doi.org/10.1088/1475-7516/2004/04/001}{\emph{JCAP}
  {\bfseries 04} (2004) 001}
  [\href{https://arxiv.org/abs/hep-th/0312100}{{\ttfamily hep-th/0312100}}].

\bibitem{Jazayeri:2016jav}
S.~Jazayeri, S.~Mukohyama, R.~Saitou and Y.~Watanabe, \emph{{Ghost inflation
  and de Sitter entropy}},
  \href{https://doi.org/10.1088/1475-7516/2016/08/002}{\emph{JCAP} {\bfseries
  08} (2016) 002} [\href{https://arxiv.org/abs/1602.06511}{{\ttfamily
  1602.06511}}].

\bibitem{Ivanov:2014yla}
M.M.~Ivanov and S.~Sibiryakov, \emph{{UV-extending Ghost Inflation}},
  \href{https://doi.org/10.1088/1475-7516/2014/05/045}{\emph{JCAP} {\bfseries
  05} (2014) 045} [\href{https://arxiv.org/abs/1402.4964}{{\ttfamily
  1402.4964}}].

\bibitem{Brennan:2017rbf}
T.D.~Brennan, F.~Carta and C.~Vafa, \emph{{The String Landscape, the Swampland,
  and the Missing Corner}},
  \href{https://doi.org/10.22323/1.305.0015}{\emph{PoS} {\bfseries TASI2017}
  (2017) 015} [\href{https://arxiv.org/abs/1711.00864}{{\ttfamily
  1711.00864}}].

\bibitem{Palti:2019pca}
E.~Palti, \emph{{The Swampland: Introduction and Review}},
  \href{https://doi.org/10.1002/prop.201900037}{\emph{Fortsch. Phys.}
  {\bfseries 67} (2019) 1900037}
  [\href{https://arxiv.org/abs/1903.06239}{{\ttfamily 1903.06239}}].

\bibitem{Bedroya:2019tba}
A.~Bedroya, R.~Brandenberger, M.~Loverde and C.~Vafa, \emph{{Trans-Planckian
  Censorship and Inflationary Cosmology}},
  \href{https://doi.org/10.1103/PhysRevD.101.103502}{\emph{Phys. Rev. D}
  {\bfseries 101} (2020) 103502}
  [\href{https://arxiv.org/abs/1909.11106}{{\ttfamily 1909.11106}}].

\bibitem{Brandenberger:2021pzy}
R.~Brandenberger, \emph{{Trans-Planckian Censorship Conjecture and Early
  Universe Cosmology}},
  \href{https://doi.org/10.31526/lhep.2021.198}{\emph{LHEP} {\bfseries 2021}
  (2021) 198} [\href{https://arxiv.org/abs/2102.09641}{{\ttfamily
  2102.09641}}].

\end{thebibliography}\endgroup

\end{document}